# Free-Standing Two-Dimensional Single-Crystalline InSb Nanosheets

D. Pan[1], D. X. Fan[2], N. Kang[2], J. H. Zhi[2], X. Z. Yu[1], H. Q. Xu[2]* and J. H. Zhao[1]*

[1]*State Key Laboratory of Superlattices and Microstructures, Institute of Semiconductors, Chinese Academy of Sciences, P.O. Box 912, Beijing 100083, China*

[2]*Key Laboratory for the Physics and Chemistry of Nanodevices and Department of Electronics, Peking University, Beijing 100871, China*

(Dated: November 21, 2015)

**Growth of high-quality single-crystalline InSb layers remains challenging in material science. Such layered InSb materials are highly desired for searching for and manipulation of Majorana fermions in solid state, a fundamental research task in physics today, and for development of novel high-speed nanoelectronic and infrared optoelectronic devices. Here we report on a new route towards growth of single-crystalline, layered InSb materials. We demonstrate the successful growth of free-standing, two-dimensional InSb nanosheets on one-dimensional InAs nanowires by molecular-beam epitaxy. The grown InSb nanosheets are pure zinc-blende single crystals. The length and width of the InSb nanosheets are up to several micrometers and the thickness is down to ~10 nm. The InSb nanosheets show a clear ambipolar behavior and a high electron mobility. Our work will open up new technology routes towards the development of InSb-based devices for applications in nanoelectronics, optoelectronics and quantum electronics, and for study of fundamental physical phenomena.**

*To whom correspondence should be addressed. E-mail: jhzhao@red.semi.ac.cn (J.H.Z.); hqxu@pku.edu.cn (H.Q.X.)

Over the past several decades, the inherent scaling limitations of Si electron devices have fuelled the exploration of alternative semiconductors, with high carrier mobility, to further enhance device performance[1-3]. In particular, high mobility III-V compound semiconductors have been actively studied[4,5]. As a technologically important III-V semiconductor, InSb is the most desired material system for applications in high-speed, low-power electronics and infrared optoelectronics owing to its highest electron mobility and narrowest bandgap among all the III-V semiconductors. Recently, epitaxially grown InSb nanostructures have been widely anticipated to have potential applications in spintronics, topological quantum computing, and detection and manipulation of Majorana fermions, due to small effective mass, strong spin-orbit interaction and giant *g* factor in InSb[6-18]. All these applications require a high degree of InSb growth control on its morphology and especially crystal quality[19,20]. Unfortunately, due to the intrinsic largest lattice parameter of InSb among all the III-V semiconductors, epitaxial growth of InSb layers faces an inevitable difficulty in finding a lattice-matched substrate. Conventionally, buffer layers with graded or abrupt composition profile are deposited on lattice mismatched substrates to obtain a layer with a required value of lattice constant[21,22]. Nevertheless, even when the sophisticated buffer-layer engineering is used, the density of dislocations threading to the surface of the buffer from its interface with a lattice mismatched substrate is often too high to grow a high crystal quality InSb layer for fabrication of high-performance nanoelectronics and quantum devices and for study of novel physical phenomena.

Here, we report on the successful growth of novel free-standing high-quality two-dimensional (2D) InSb nanosheets by molecular-beam epitaxy (MBE). A new route of growing high material quality layered InSb structures is discovered, in which free-standing InSb nanosheets are epitaxially grown on InAs nanowire stems and thus the process is independent of buffer-layer engineering. The morphology and size of free-standing InSb nanosheets can be controlled in the approach by tailoring the Sb/In beam equivalent pressure (BEP) ratio and InSb growth time. We demonstrate the growth of free-standing InSb nanosheets with the length and width up to several micrometers and the thickness down to ~10 nm. High-resolution transmission electron microscope (TEM) images show that the grown InSb nanosheets are pure zinc-blende (ZB) single crystals and have excellent epitaxial relationships with the InAs nanowire stems. The formation of the InSb nanosheets is attributed to a combination of vapor-liquid-solid (VLS) and anisotropic lateral growth. Electrical measurements show that the grown InSb nanosheets exhibit an ambipolar behavior and a high electron mobility. These novel, high material quality, free-standing InSb single-crystalline nanosheets have the great potential not only for applications in high-speed electronics and infrared optoelectronics, but also for realization of novel quantum devices for the studies of fundamental physics.

The 2D InSb nanosheets were grown by MBE on free-standing InAs nanowire stems which were first grown on Si (111) substrates using Ag as catalyst in the MBE chamber[23]. Figure 1a shows the schematics for the growth process. We found that the morphology of InSb strongly depends on the Sb/In BEP ratio, and the InSb nanosheets can be realized by tailoring the Sb/In BEP ratio (Supplementary Section S1). For the sample grown with low Sb/In BEP ratio of 1~20, InSb and InAs formed core-shell or axial heterostructure nanowires (Supplementary Figs. S1 and S2). Further increasing the Sb/In BEP ratio, the resulting InSb nanowires have diameters obviously larger than that of the InAs segment (Detailed TEM investigation of a dozen of such nanowires reveals a diameter increase from 130% to 589%, see Supplementary Fig. S3 and Table S1). By increasing the Sb/In BEP ratio to the range of 27-80, new geometrically structured materials with each consisting of a 2D InSb nanosheet and a 1D InAs nanowire stem were obtained (Supplementary Figs. S1 and S4). Figures 1b to 1g show the top-view magnified scanning electron microscope (SEM) images of InSb nanosheets grown with an Sb/In BEP ratio of 80. As can be seen, the grown InSb nanosheets have parallelogram shapes. The thicknesses of the InSb nanosheets show significant variation (Supplementary Fig. S5), as roughly measured from SEM images, from ~67 nm (Fig. 1b), to ~30 nm (Figs. 1c and 1d), and to the ultrathin value of ~10 nm (Figs. 1e to 1g). The ultrathin nanosheets can be penetrated easily by the electron beam of the SEM (In Figs. 1e to 1g, features on the substrate surface are clearly visible through the thin InSb nanosheets). Figures 1h and 1i show the side-view magnified SEM images of InSb nanosheets in the same sample and it is clear that the InSb nanosheets can be grown on ultrathin InAs nanowires (~10 nm in diameter) which oriented perpendicular to the substrate surface with a pure wurtzite (WZ) crystal structure[23].

To examine the structural characteristics, crystalline quality and the chemical composition of the grown InSb nanosheets, TEM and energy dispersive x-ray spectroscopy (EDS) measurements were performed. Figure 2a is a bright-field TEM image of a typical InSb nanosheet grown on an InAs nanowire at an Sb/In BEP ratio of 27. The InAs stem is 21 nm in diameter and 725 nm in length, while the InSb segment has a parallelogram shape with side lengths of 380 nm and 508 nm. It is found that the 2D InSb nanosheets can be successfully grown not only on 1D WZ crystalline InAs nanowires (Supplementary Fig. S6) but also on 1D InAs nanowires with ZB phase, as shown in Fig. 2b. High-resolution TEM images of the side sections (Figs. 2c, 2f and 2h), the corner sections (Figs. 2d and 2g) and the section near the tip (Fig. 2e) of the InSb nanosheet and the associated Fourier transform (Fig. 2i) illustrate that the InSb nanosheet has a perfect ZB crystal structure, free from stacking faults or WZ regions. Although the stacking faults have been observed in Ag-catalyzed and self-seeded InSb nanowires by other groups[24,25], detailed TEM observations of our grown InSb nanosheets with different shapes and sizes all reveal that the InSb nanosheets are fully single-crystalline, completely free from stacking faults and twinning defects (Supplementary Figs. S7 and S8). As observed in InSb nanocrystals grown with a high Sb/In BEP ratio[26], the side facets with low surface energy such as $\{\bar{1}11\}$ and $\{011\}$ can be clearly seen in our InSb nanosheets (Fig. 2d, Supplementary Fig. S7). High-angle annular dark-field scanning TEM (HAADF-STEM) and corresponding EDS line profiles (Supplementary Fig. S9) show that the InAs/InSb heterostructures start as InAs and then change to InSb. EDS elemental mappings shown in Figs. 2j to 2m confirm that sharp interfaces formed between the InAs nanowires and the InSb nanosheets as reported in InAs/InSb heterostructure nanowires[27-30]. The remaining spherical catalyst particle on the top of InSb nanosheets is found to be composed of Ag, In and Sb (Fig. 2n). EDS line point analysis indicates that the InSb nanosheet contains In and Sb with an atomic ratio of ~1:1 and the contents of Ag, In and Sb in the seed particles after the growth of the InSb nanosheets and of the InSb nanowires are similar (Supplementary Table S2).

The length and width of the InSb nanosheets can be controlled directly by tailoring the InSb growth time. Figures 3a to 3d show the SEM images of the InSb nanosheets obtained with the InSb growth time of 40 min. Comparing to the InSb nanosheets obtained with the InSb growth time of 80 min (Fig. 1, Supplementary Fig. S1), no apparent difference in morphology is observed in these InSb nanosheets, except for their smaller sizes. With increasing the InSb growth time to 120 min and to 160 min (Figs. 3e to 3l) while keeping other growth parameters unchanged, the obtained InSb nanosheets are still in planar shapes—these InSb nanosheets can be grown to micrometers in length and width, while still be kept in a thin thickness.

As to the growth mechanism of the InSb nanosheets, it is unlikely that the growth is governed solely by the traditional VLS[31,32] or vapor-solid (VS) process[33]. On the contrary, we consider that the growth of the InSb nanosheets is dominated by the combination of VLS (vertical growth) and VS (lateral growth) processes. The spherical Ag-In-Sb alloy particles on the top of InSb nanowires and InSb nanosheets have the same crystal structure and the similar compositions (Supplementary Fig. S10 and Table S2), indicating that the VLS mechanism for Ag-seed InSb nanowire growth[24] exists in the growth of the InSb nanosheets. Meanwhile, we find that the lateral growth observed often in the growth of antimonide nanowires[27-30] is incorporated into our InSb nonsheet growth, since the width of the nanosheets is growth time dependent (Figs. 3a to 3l). However, the lateral growth for the nanosheets is quite different from that for nanowires. The InSb nanowires share a rotationally symmetric, laterally overgrown shell[34], while the InSb nanosheets show an anisotropic lateral crystal growth. One possible combined VLS and anisotropic lateral growth process of the InSb nanosheets is given in Supplementary Section S6. Although the exact reason for the anisotropic lateral growth of the InSb nanosheets remains to be determined, we believe that the mechanism of combination of the VLS and the anisotropic lateral growth could be used to fabricate other high-quality 2D antimonide nanostructures.

Electronic properties of the grown InSb nanosheets were characterized by electrical measurements. Figure 4a is a tilted-view sketch of the device structure used in the characterization (lower panel) and a corresponding atomic force microscopy image of a typical fabricated device (upper panel). In the device, an InSb nanosheet was contacted by Ti/Au electrodes in a Hall-bar configuration and the carrier density in the nanosheet was modulated by a global back gate. Details about the device fabrication and measurement scheme can be found in Methods and Supplementary Section S7. First, by voltage-biasing the source (S) and drain (D) contacts as shown in Fig. 4a, the 2-probe conductance $G$, which takes the form $G = I_{ds}/V_{ds}$, was measured as a function of gate voltage $V_{gs}$ at different temperatures (Fig. 4b, main panel, Device 1). When lowering temperature, several characteristic features were observed. (1) The off state $G$ shows a monotonic decrease for $V_{gs}$ in the region of -3 V to 0 V, while the on state $G$ shows a monotonic increase for $V_{gs}$ in the region of 8 V to 10 V. (2) The $G$-$V_{gs}$ curve becomes steeper in the linear region, indicating an increase of the peak transconductance $g_m$, where $g_m$ = max$\{dG/dV_{gs}\}$. (3) A clear ambipolar transport characteristic is seen at $T$ = 60 mK, with the conductance on the hole side 1-2 orders of magnitude smaller than the electron side (Fig. 4b, inset). At this temperature, $G$ can be gate-tuned from $6G_0$ at on state down to $0.008G_0$ at off state ($G_0 = 2e^2/h$, $e$ is the elementary charge, and $h$ is the Planck constant). Moreover, the $G$-$V_{gs}$ curves are well reproduced for the upward (blue) and downward (red) back-gate-voltage sweep directions, indicating a good surface quality of our InSb nanosheets, free from major influence of interfacial charge traps in spite of its large surface area. Figures 4c and 4d show the $I_{ds}$-$V_{ds}$ plots at various gate voltages in an n- and a p-type region, respectively. The linear $I_{ds}$-$V_{ds}$ curves obtained in the n-type region indicate an ohmic behavior in the electron injection and the absence of Schottky contact barrier. The nonlinear $I_{ds}$-$V_{ds}$ curves observed in the p-type region indicate, on the contrary, the presence of injection barrier for holes.

To obtain a reliable field-effect mobility, especially at low temperatures, one needs to take into account the contact resistance. Here we adopt a pinch-off trace fitting method[19,29,35]. Yet, a major drawback remains in the approach due to the uncertainty in obtaining the gate-to-nanosheet capacitance $C_g$. To circumvent it, the Hall-bar device structure was employed to experimentally extract $C_g$, although $C_g$ could be estimated from the geometric size of the device. Figure 5 summarizes the low-field Hall measurement data obtained from a second device (Device 2). The Hall resistance, $R_{xy}$, is gate-tunable and approaches ~2 kΩ at $B \approx 1$ T for the low $V_{gs}$ region (Fig. 5a, main panel). A linear fit to the $R_{xy}$-$B$ curve yields the Hall coefficient (Fig. 5a, inset). The carrier density $n$ is then extracted from the gate-dependent Hall coefficient and is found to increase linearly from ~$2\times10^{11}$ cm$^{-2}$ to ~$1.6\times10^{12}$ cm$^{-2}$ with increasing $V_{gs}$ (Fig. 5b, main panel). Then, a gate-to-nanosheet capacitance $C_g$ = 832 μF is determined from the fitting slope of the $n$-$V_{gs}$ curve, close to the value of ~800 μF estimated based on the geometry of the device. This measured value is used to fit the 2-probe $G$-$V_{gs}$ curve to evaluate the field-effect mobility of Device 2 (Fig. 5b, inset). The analysis yields an electron mobility of ~15,000 cm$^2$ V$^{-1}$ s$^{-1}$ for this device. An electron mobility of ~18,500 cm$^2$ V$^{-1}$ s$^{-1}$ is obtained for Device 1 by the same kind of analysis (Supplementary Section S8).

In conclusion, we demonstrate a new growth route of high-quality 2D InSb layers by MBE. These InSb layers are free-standing 2D InSb nanosheets grown on 1D InAs nanowires, which is independent of conventional buffer-layer engineering. The morphology and size of InSb nanosheets can be controlled by tailoring the Sb/In BEP ratio and growth time. The length and width of the grown InSb nanosheets can be up to several micrometers and the thickness can be down to ~10 nm. The InSb nanosheets are pure ZB single crystals. The electrical measurements show that these InSb nanosheets exhibit a high electron mobility and an ambipolar behavior. Our work opens a conceptually new approach to obtaining high-quality 2D narrow bandgap semiconductor nanostructures, and will speed up the applications of InSb nanostructures in nanoelectronics, optoelectronics and quantum electronics and in the development of topological quantum computation technologies.

**Methods**

The InAs/InSb nanostructures were grown in a solid source molecular-beam epitaxy (MBE, VG80) system. Commercial p-type Si (111) wafers were used as the substrates. Before loading the Si substrates into the MBE chamber, they were immersed in a diluted HF (2%) solution for 1 min to remove the surface contamination and native oxide. After cleaning, a Ag layer of 2 Å nominal thickness was deposited on the substrate in the MBE growth chamber at room temperature and then annealed *in situ* at 650 $^0$C for 20 min to generate Ag nano-particles. InAs nanowires were grown for 20 min at a temperature of 505 $^0$C with an As/In beam equivalent pressure (BEP) ratio of 30. Then the group-V source was abruptly switched from As to Sb without any variation of substrate temperature. All the InSb segments (if no specific description) were grown for 80 min at different Sb/In BEP ratios by increasing the Sb flux while keeping the In flux constant.

The morphologies of the samples were observed by scanning electron microscope with a Nova NanoSEM 650, operated at 20 kV. Structural characterization was performed using transmission electron microscope (TEM) and samples were removed from the growth substrate via sonication in ethanol and then drop-cast onto a holey carbon film supported by a copper grid. High-resolution TEM images and energy-dispersive x-ray spectrometer (EDS) spectra (including the EDS elemental mapping and line scans shown in Fig. 2) were taken with the JEM-ARM200F, operated at 200 kV. Other high resolution TEM images shown in Figs. S2, S3, S6, S8 and S10 were collected using a JEOL-2011, operated at 200 kV. Selected area electron diffraction patterns in Fig. S7 were acquired using a TECNAI G$^2$20, operated at 200 kV. The chemical composition shown in Fig. S9 was evaluated by scanning TEM, using an FEI Titan, operated at 300 kV and in both the line scan and point analysis modes. Atomic force microscopy measurements were carried out with a Digital Instruments Nanoscope IIIA using silicon tips with a typical resonance frequency of 300 kHz.

For device fabrication, the as-grown InSb nanosheets were first mechanically transferred to a degenerately n-doped Si substrate covered by 105 nm SiO$_2$ layer, which serves as a global back gate. Then selected nanosheets were optically positioned relative to predefined alignment marks. Finally, the contact electrodes were patterned onto the sample using standard electron-beam lithography process. Prior to electron-beam evaporating a layer of 10/100 nm Ti/Au metal film, the sample was chemically etched for 2 min in a DI water-diluted (NH$_4$)$_2$S$_x$ solution to remove the surface oxide layer at the contact area. All electrical measurements in this work were performed in a $^3$He/$^4$He dilution fridge with a base temperature of 8 mK (Oxford Triton 200). In the 2-probe measurements, only the source/drain (S/D) contacts were dc voltage-biased, and the other four contacts were left floated. In the Hall-bar measurements, the S/D contacts were used to inject a constant dc current bias of ~50 nA through the sample, and the Hall voltage $V_{xy}$ was amplified and recorded simultaneously with respect to the sweep of the magnetic field applied perpendicularly to the plane of the sample/substrate. In the 2-probe $G$-$V_{gs}$ data, a circuit resistance of ~21.56 kΩ (including RC filters and serial resistors) has been subtracted.

**References**

1. Lundstrom, M. Moore's Law Forever? *Science* **299**, 210-211 (2003).
2. Chau, R., Doyle, B., Datta, S., Kavalieros, J., Zhang, K. Integrated nanoelectronics for the future. *Nature Mater.* **6**, 810-812 (2007).

3. Bryllert, T., Wernersson, L. E., Froberg, L. E., Samuelson, L. Vertical high-mobility wrap-gated InAs nanowire transistor. *IEEE Electron Device Lett.* **27,** 323-325 (2006).
4. Ko, H. *et al*. Ultrathin compound semiconductor on insulator layers for high-performance nanoscale transistors. *Nature* **468,** 286-289 (2010).
5. Tomioka, K., Yoshimura, M., Fukui, T. A III-V nanowire channel on silicon for high-performance vertical transistors. *Nature* **488,** 189-193 (2012).
6. Wolf, S. A. *et al*. Spintronics: A spin-based electronics vision for the future. *Science* **294,** 1488-1495 (2001).
7. Zutic, I., Das Sarma, S. Spintronics: Fundamentals and applications. *Rev. Mod. Phys.* **76,** 323- 410 (2004).
8. Sau, J. D., Tewari, S., Das Sarma, S. Generic new platform for topological quantum computation using semiconductor heterostructures. *Phys. Rev. Lett.* **104,** 040502-040505 (2010).
9. Alicea, J. Majorana fermions in a tunable semiconductor device. *Phys. Rev. B* **81,** 125318-1253127 (2010).
10. Mourik, V. *et al*. Signatures of majorana fermions in hybrid superconductor-semiconductor nanowire devices. *Science* **336,** 1003-1007 (2012).
11. Deng M. T. *et al*. Anomalous zero-bias conductance peak in a Nb-InSb nanowire-Nb hybrid device. *Nano Lett.* **12,** 6414-6419 (2012).
12. Deng M. T. *et al*. Parity independence of the zero-bias conductance peak in a nanowire based topological superconductor-quantum dot hybrid device. *Sci. Rep.* **4,** 7261-7268 (2014).
13. Rokhinson, L. P., Liu, X. Y., Furdyna, J. K. The fractional a.c. Josephson effect in a semiconductor-superconductor nanowire as a signature of Majorana particles. *Nature Phys.* **8,** 795-799 (2012).
14. Vurgaftman, I., Meyer, J. R., Ram-Mohan, L. R. Band parameters for III-V compound semiconductors and their alloys. *J. Appl. Phys.* **89,** 5815-5875 (2001).
15. Nilsson, H. A. *et al*. Giant, level-dependent *g* factors in InSb nanowire quantum dots. *Nano Lett.* **9,** 3151-3156 (2009).
16. Nilsson, H. A. *et al*. Correlation-induced conductance suppression at level degeneracy in a quantum dot. *Phys. Rev. Lett.* **10,** 186804-186807 (2010).
17. Nilsson, H. A., Samuelsson, P., Caroff, P., Xu, H. Q. Supercurrent and multiple Andreev reflections in an InSb nanowire Josephson junction. *Nano Lett.* **12,** 228-233 (2012).
18. Fan, D. *et al*. Formation of long single quantum dots in high quality InSb nanowires grown by molecular beam epitaxy. *Nanoscale* **7,** 14822-14828 (2015).
19. Plissard, S. R. *et al*. Formation and electronic properties of InSb nanocrosses. *Nature Nanotech.* **8,** 859-864 (2013).
20. Car, D. *et al*. Rationally designed single-crystalline nanowire networks. *Adv. Mater.* **26,** 4875-4879 (2014).
21. Weng, X. *et al*. Effects of buffer layers on the structural and electronic properties of InSb films. *J. Appl. Phys.* **97,** 043713-043719 (2005).
22. Mishima, T. D., Edirisooriya, M., Goel, N., Santos, M. B. Dislocation filtering by $Al_xIn_{1-x}Sb$ / $Al_yIn_{1-y}Sb$ interfaces for InSb-based devices grown on GaAs (001) substrates. *Appl. Phys. Lett.* **88,** 191908-191910 (2006).
23. Pan, D. *et al*. Controlled synthesis of phase-pure InAs nanowires on Si(111) by diminishing the diameter to 10 nm. *Nano Lett.* **14,** 1214-1220 (2014).
24. Vogel, A. T. *et al*. Ag-assisted CBE growth of ordered InSb nanowire arrays. *Nanotechnology* **22,** 015605-015610 (2011).
25. Mandl, B. *et al*. Crystal structure control in Au-free self-seeded InSb wire growth. *Nanotechnology* **22,** 145603-145609 (2011).
26. Lin, A. Shapiro, J. N., Eisele, H., Huffaker, D. L. Tuning the Au-free InSb nanocrystal morphologies grown by patterned metal-organic chemical vapor deposition. *Adv. Funct. Mater.* **24,** 4311-4316 (2014).
27. Caroff, P. *et al*. High-quality InAs/InSb nanowire heterostructures grown by metal-organic vapor-phase epitaxy. *Small* **4,** 878-882 (2008).
28. Ercolani, D. *et al*. InAs/InSb nanowire heterostructures grown by chemical beam epitaxy. *Nanotechnology* **20,** 505605-505610 (2009).
29. Plissard, S. R. *et al*. From InSb nanowires to nanocubes: looking for the sweet spot. *Nano Lett.* **12,** 1794-1798 (2012).
30. de la Mata, M., Magen, C., Caroff, P., Arbiol, J. Atomic scale strain relaxation in axial semiconductor III-V nanowire heterostructures. *Nano Lett.* **14,** 6614-6620 (2014).
31. Wagner, R. S., Ellis, W. C. Vapor-liquid-solid mechanism of single crystal growth. *Appl. Phys. Lett.* **4,** 89-90 (1964).
32. Morales, A. M., Lieber, C. M. A laser ablation method for the synthesis of crystalline semiconductor nanowires. *Science* **279,** 208-211 (1998).
33. Pan, Z. W., Dai, Z. R., Wang, Z. L. Nanobelts of semiconducting oxides. *Science* **291,** 1947-1949 (2001).
34. Ercolani, D. *et al*. Growth of InAs/InAsSb heterostructured nanowires. *Nanotechnology* **23,** 115606-115614 ( 2012).
35. Gül, Ö. *et al*. Towards high mobility InSb nanowire devices. *Nanotechnology* **26,** 215202-215208 (2015).

## Acknowledgments
The authors acknowledge X. A. Yang at Institute of Physics, Chinese Academy of Sciences, and M. F. Wang at Tsinghua University for their assistance with TEM measurements. This work was supported by the MOST of China (Grant Nos. 2012CB932701 and 2012CB932703) and the National Natural Science Foundation of China (Grant Nos. 61504133, 91221202, 91421303 and 61321001).

**Additional information**

Supplementary Materials
Figs. S1 to S15
Tables S1 to S3
References

**Competing financial Interests**

The authors declare that they have no competing financial interests.

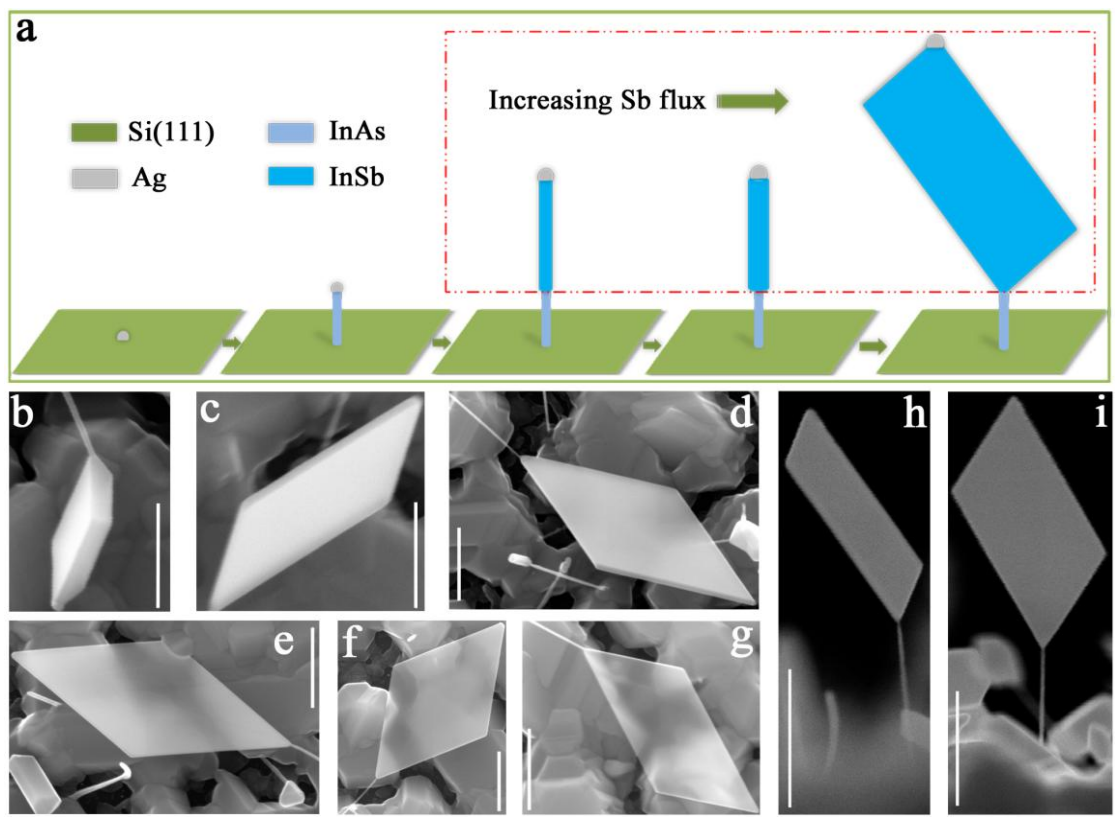

**Figure 1 | Diagram of the growth process and SEM images of the InSb nanosheets**. **a**, Schematic demonstration of the growth process of the 2D InSb nanosheets. By tailoring the Sb/In BEP ratio, the 2D InSb nanosheets were epitaxially grown on free-standing InAs nanowire stems which were first grown on Si (111) substrates using Ag as catalyst. **b-g**, **h** and **i**, Top view (**b-g**) and side view (**h** and **i**) SEM images of the InSb nanosheet epitaxially grown on 1D InAs nanowires with an Sb/In BEP ratio of 80. The scale bars in **b** and **c** are 400 nm. The scale bars in **d** to **i** are 500 nm.

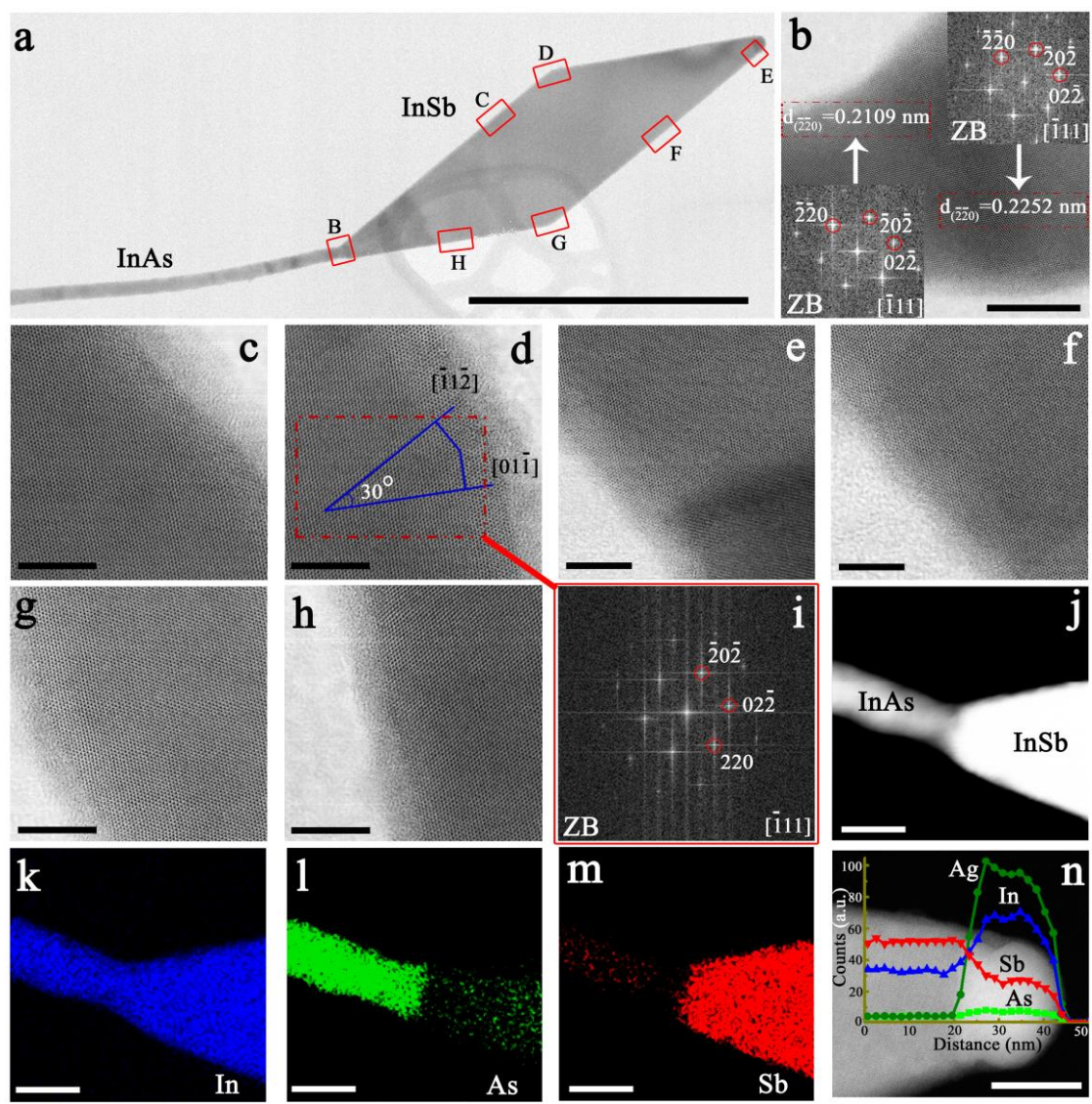

**Figure 2 | Crystal structure and chemical composition of an InSb nanosheet. a**, Low-resolution TEM image of an InAs/InSb nanowire-nanosheet grown with an Sb/In BEP ratio of 27. The red rectangles highlight the regions where the high-resolution TEM images were recorded. The scale bar in the figure is 500 nm. **b**, High-resolution TEM image taken from the InAs/InSb interface (region B in **a**). Here the scale bar is 10 nm. The left and right insets are the corresponding FFTs of the InAs and InSb segments, respectively. The interplanar spacings of the InAs and InSb segments were measured using a standard digital micrograph software. **c-h**, High-resolution TEM images taken from the regions C, D, E, F, G and H in **a**, respectively. Here the scale bars are 5 nm. **i**, FFT pattern corresponding to **d**. **j-m**, HAADF-STEM image of the InAs/InSb interface area (**j**) and false-color STEM EDS elemental maps (**k** to **m**) of In, As, and Sb in the region of the InAs/InSb junction. Here the scale bars are 25 nm. **n**, HAADF-STEM image of the InSb nanosheet in the region covering the InSb/seed-particle interface with overlaid EDS profiles. The scale bar is 20 nm in the figure.

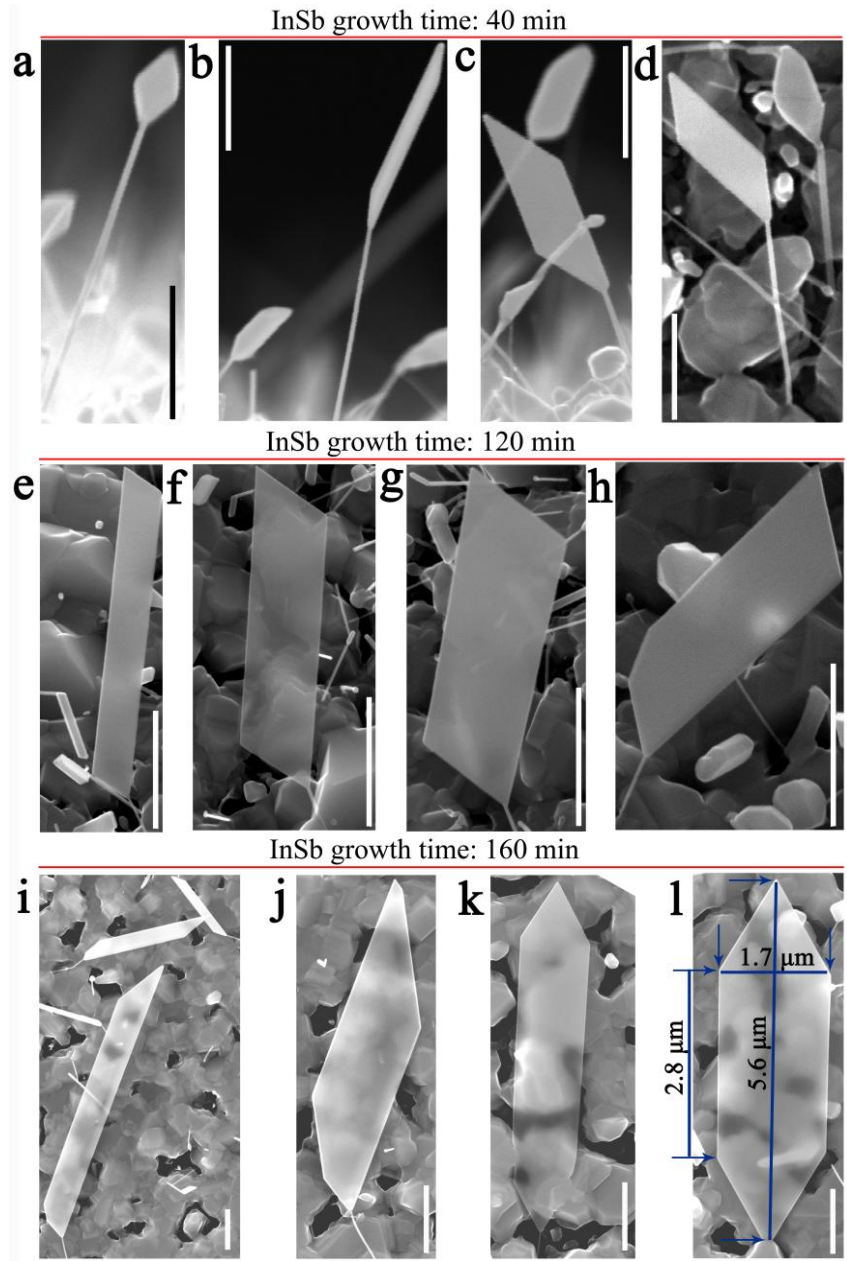

**Figure 3 | Size controlling of the InSb nanosheets by tailoring the InSb growth time. a-c**, **d**, Side view (**a-c**) and top view (**d**) SEM images of the InSb nanosheets with the growth time of 40 min. The scale bars are 500 nm in these figures. **e-g**, **h**, 20 ° tilted view (**e-g**) and top view (**h**) SEM images of the InSb nanosheets with the growth time of 120 min. Here, the scale bars are 1 μm. **i-l**, Top view SEM images of the InSb nanosheets with the growth time of 160 min. Here the scale bars are 1 μm.

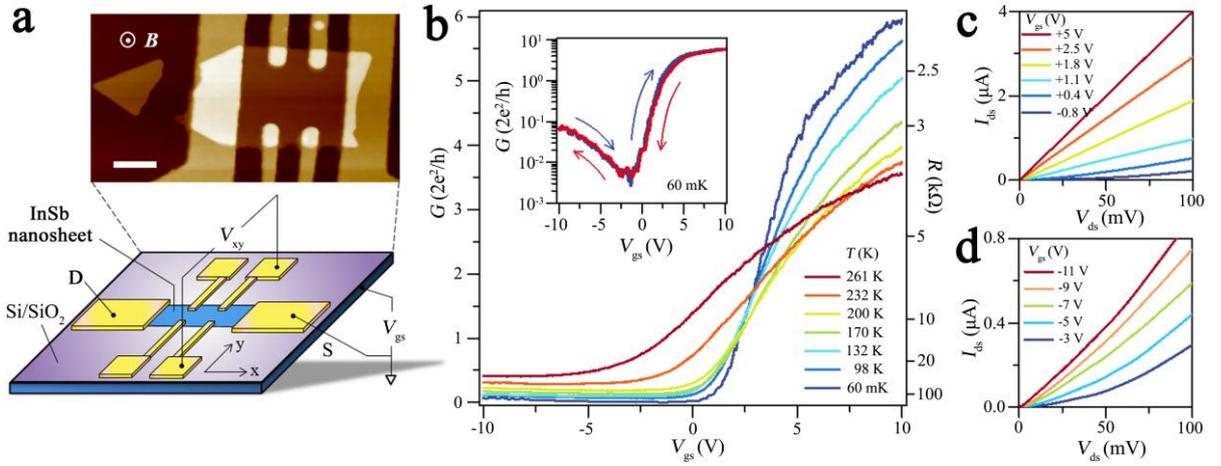

**Figure 4 | Transport properties of the InSb nanosheets. a**, Schematic diagram (lower panel) and atomic force microscopy image (upper panel) of typical InSb nanosheet Hall-bar device. Here, the scale bar is 500 nm. A voltage bias is applied to the source (S) and drain (D) electrode in a 2-probe measurement setup, while a current bias is applied to the S and D electrode in the Hall measurement setup. A global back gate voltage $V_{gs}$ is applied to the degenerately doped Si substrate. The magnetic field is perpendicular to the plane of the sample. **b**, 2-probe conductance $G$ as a function of $V_{gs}$ at different temperatures (Device 1). Here, in the measurements, the S-D bias $V_{ds}$ applied is between 10 mV (near pinch-off) and ~1 mV (at $V_{gs}$= 10 V). Note that the right axis shows the 2-probe resistance $R = 1/G$. The inset shows a logarithmic scale plot of the $G$-$V_{gs}$ curve at $T = 60$ mK. Both the upward (blue) and downward (red) sweep show a clear ambipolar characteristic and a small hysteresis (of ~0.8 V). **c**, Corresponding current-voltage characteristics at different back gate voltages in the electron transport region. **d**, Corresponding current-voltage characteristics at different back gate voltages in the hole transport region.

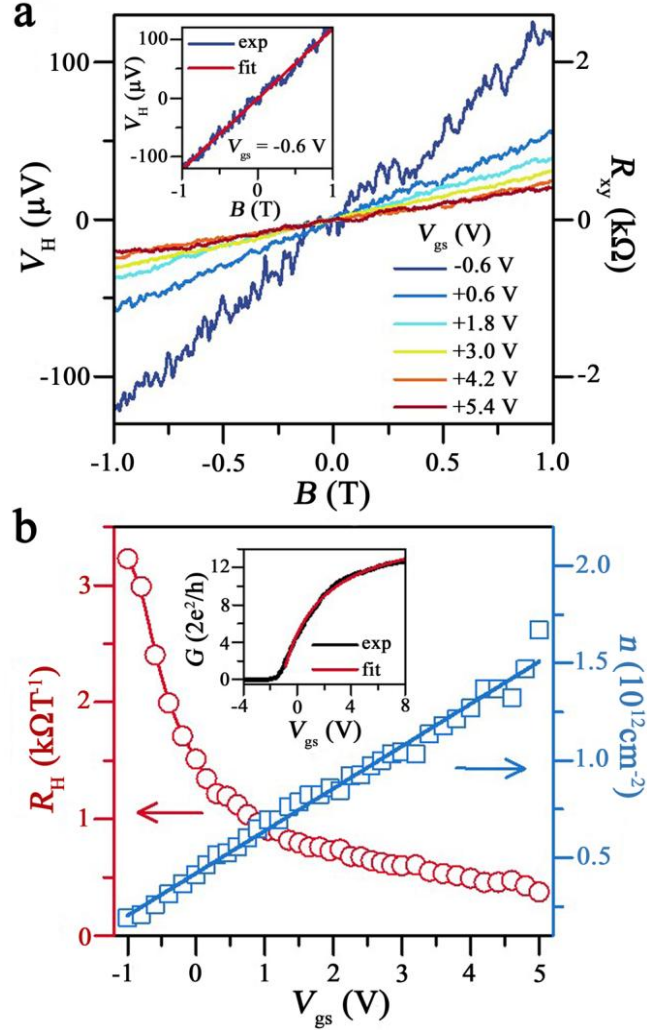

**Figure 5 | Hall measurements of an InSb nanosheet device. a**, Hall voltage $V_H$ measured as a function of magnetic field $B$ at a constant source-drain current bias of ~50 nA and $T$ = 60 mK for several gate voltages (Device 2). Here the right axis shows the transverse resistance $R_{xy}$ of the form $R_{xy} = V_H/I$. Note that larger Hall resistance values are found for lower gate voltages (lower carrier densities). The inset shows a linear fit (red) to the measured $V_H$-$B$ trace (blue), from which the Hall coefficient $R_H$ can be determined. **b**, Hall coefficient $R_H$ obtained as a function of $V_{gs}$ (left axis, circle) at $T$ = 60 mK. From $R_H$, we can extract the sheet carrier density $n$ as a function of $V_{gs}$ (right axis, square). A linear fit to the $n$-$V_{gs}$ data yields $dn/dV_{gs}$ = 2.1×10$^{11}$ cm$^{-2}$ V$^{-1}$ and therefore a gate-to-nanosheet capacitance $C_g$ = 832 μF. Using this $C_g$ value, we can fit the measured $G$-$V_{gs}$ trace of the device (inset of Fig. 5b) and extract a field-effect mobility of ~15,000 cm$^2$ V$^{-1}$ s$^{-1}$.

# Supplementary information

## Free-Standing Two-Dimensional Single-Crystalline InSb Nanosheets

D. Pan[1], D. X. Fan[2], N. Kang[2], J. H. Zhi[2], X. Z. Yu[1], H. Q. Xu[2]* and J. H. Zhao[1]*

[1]*State Key Laboratory of Superlattices and Microstructures, Institute of Semiconductors, Chinese Academy of Sciences, P.O. Box 912, Beijing 100083, China*

[2]*Key Laboratory for the Physics and Chemistry of Nanodevices and Department of Electronics, Peking University, Beijing 100871, China*

*To whom correspondence should be addressed. E-mail: jhzhao@red.semi.ac.cn (J.H.Z.); hqxu@pku.edu.cn (H.Q.X.)

**Contents**

**S1  Morphology controlling of the InSb nanostructures**

**S2  Thickness information of the InSb nanosheets**

**S3  Detailed crystal structure and quality information of the InSb nanosheets**

**S3.1 TEM images of an InSb nanosheet grown on a WZ InAs nanowire**

**S3.2 Selective area electron diffraction patterns of an InSb nanosheet**

**S3.3 TEM images of a large size InSb nanosheet**

**S4  Chemical composition of the InSb nanosheets**

**S5  Seed-particles information of the InSb nanostructures**

**S6  A nucleation process of the InSb nanosheets**

**S7  Summary of the Hall-bar device parameters**

**S8  Extraction of the field-effect mobility in a nanosheet**

**S8.1 Hall measurement on Dev-1**

**S8.2 Temperature-dependent field-effect motilities of Dev-1 and Dev-2**

## S1  Morphology controlling of the InSb nanostructures

We find that the morphology of the InSb nanostructures strongly depends on the Sb/In beam equivalent pressure (BEP) ratio, and the InSb nanosheets can be realized by tailoring the Sb/In BEP ratio. The scanning electron microscope (SEM) images in Fig. S1 show the 20°-tilted view of the InAs/InSb heterostructures grown at different Sb/In BEP ratios. For the sample with an Sb/In BEP ratio of 1 (Fig. S1(a)), the nanowires exhibit tapered morphologies with short lengths since the axial growth rate of InSb is limited by the low Sb flux and an InSb shell forms around the top of the InAs nanowire stem, instead of an axial heterostructure as reported in other antimonide nanowires[1]. It is noteworthy that this low Sb/In BEP ratio is very close to typical Sb/In BEP ratio for the molecular-beam epitaxy (MBE) growth of InSb planar epitaxial layers[2]. Therefore, a homogeneous InSb layer can be observed on the substrate surface. At a slightly higher Sb/In BEP ratio of 20, InAs/InSb axial heterostructure nanowires are obtained and a typical nanowire is shown in Fig. S1(b). As can be seen, both the InAs and the InSb sections appear fully un-tapered. The diameter of InSb nanowire is small (~28 nm, Fig. S2), indicating the absence of obvious lateral growth. However, further increasing the Sb/In BEP ratio up to 27, it is clear that the resulting InSb nanowires have diameters obviously larger than that of the InAs segment (Fig. S1(c)). Detailed transmission electron microscope (TEM) observation of a dozen such nanowires revealed a diameter increase from 130% to 589% (Table S1 and Fig. S3), which is much larger than the diameter variation of the InAs/InSb heterostructure nanowires grown by metal-organic vapor phase epitaxy (MOVPE) and chemical beam epitaxy (CBE)[3,4]. Interestingly, at this Sb/In BEP ratio we observed a new geometrical structure that consists of two dimensional (2D) InSb nanosheets and one dimensional (1D) InAs nanowires. InSb nanosheets in the shapes of parallelogram, pentagon and hexagon are observed and their dimensions will be presented later. For even larger Sb/In BEP ratios (50-80, Figs. S1(d) and S1(e)), the primary *morphologies* of the InAs/InSb heterostructures are nanowire-nanosheet. It is also worth noting that the surface of the Si substrate is covered by many irregular

InSb islands when the Sb/In BEP ratios is larger than 1. Moreover, the density of the parasitic InSb islands on the substrate surface increases with increasing Sb/In BEP ratio. When the Sb/In BEP ratio is above 120 (Fig. S1(f)), no InSb nanowires or nanosheets were observed since the InAs nanowires were covered by the parasitic InSb islands.

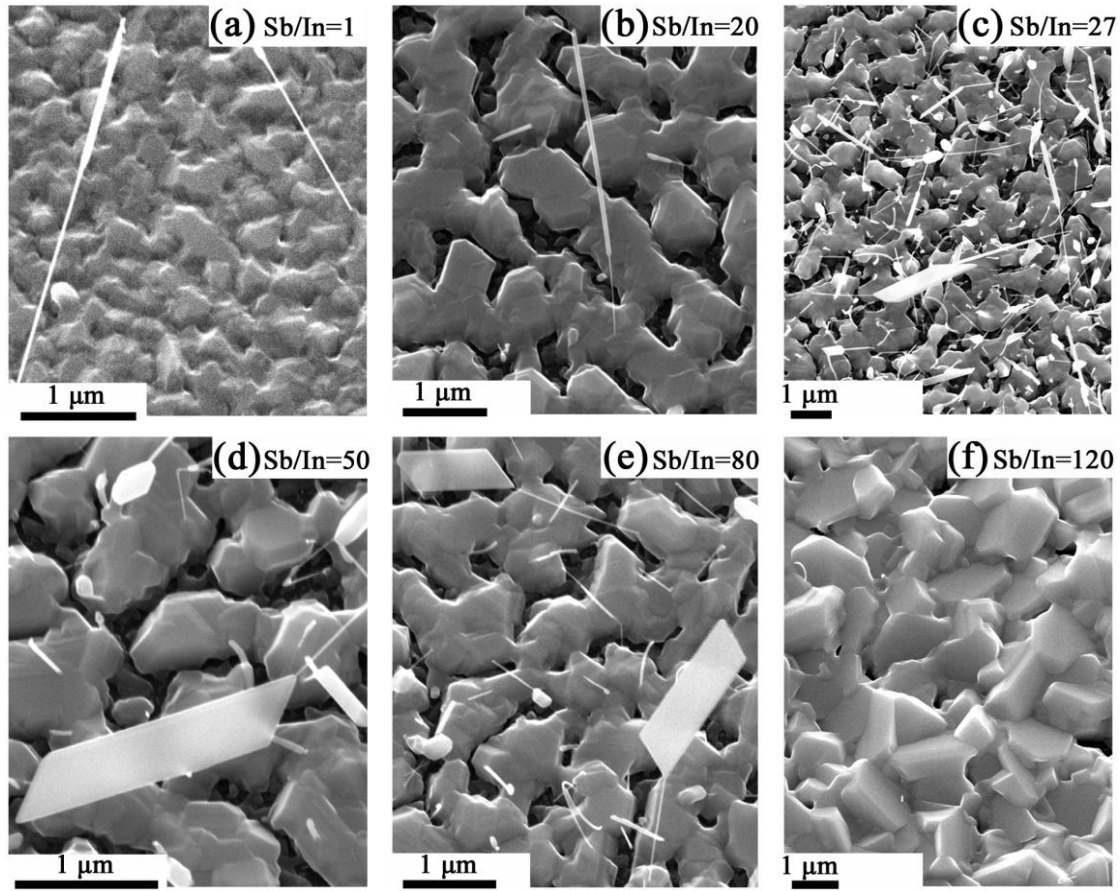

Fig. S1. SEM images (20 °tilt) of InAs/InSb nanostructures grown at different Sb/In BEP ratios: (a) Sb/In = 1; (b) Sb/In = 20; (c) Sb/In = 27; (d) Sb/In = 50; (e) Sb/In = 80; (f) Sb/In = 120.

Figure S2(a) is a bright-field TEM image of a typical InSb nanowire grown with an Sb/In BEP ratio of 20. The high-resolution TEM image recorded from the InSb region of the nanowire (Fig. S2(b)) indicates a diameter of 28 nm. Although very thin InSb nanowires down to 5 nm have been reported[5], to our knowledge, this is the first observation of ultrathin InSb nanowire with diameter down to 30 nm in a heterostructure with an InAs nanowire. Although the diameter of the InSb is larger than that of the InAs segment (diameter is about 18 nm), the radial growth of the InSb nanowires is not so obvious compared with the heterostructures grown with higher

Sb/In BEP ratios (to be shown later) and high aspect ratios up to 89 (2.5 μm in length) can be obtained.

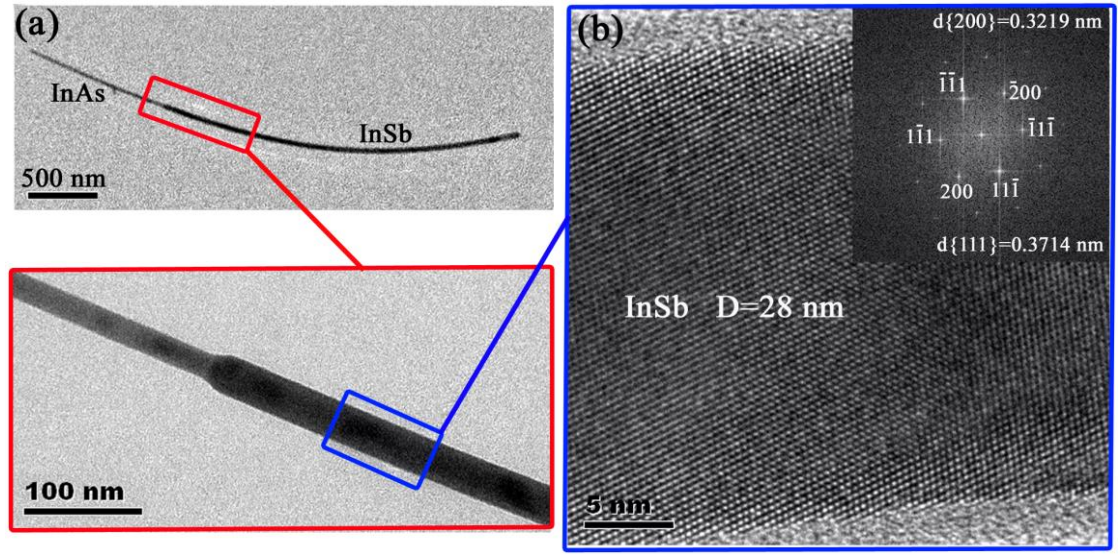

Fig. S2. TEM images of an InAs/InSb axial heterostructure nanowire grown with Sb/In = 20. The top panel of (a) shows the low magnification overview of the heterostructure nanowire. The thin and the thick portions are InAs and InSb nanowires, respectively. The bottom panel of (a) shows the corresponding magnified view of the junction region. The blue rectangle highlights the region where the high-resolution images were recorded. (b) displays high-resolution TEM image taken from InSb region. Inset: the corresponding FFT of the image.

Figure S3(a) shows a TEM image of a typical InSb nanowire grown with an Sb/In BEP ratio of 27. As can be seen, the InSb nanowire with a diameter (143 nm) about 6 times larger than that of the InAs (24 nm) can be observed (the aspect ratio is about 5). High-resolution TEM images as shown in Figs. S3(b) to S3(f) indicate that the InAs grows along the $<000\bar{1}>$ direction with a hexagonal wurtzite (WZ) phase and the InSb grows along the $<11\bar{1}>$ direction with a cubic zinc-blende (ZB) phase. It is worth noting that the lateral growth on the InSb nanowire along the $<2\bar{1}1>$ and $<01\bar{1}>$ directions can be clearly observed. Table S1 lists the diameters of the InAs, InSb and seed particle for 6 representative heterostructure nanowires grown with an Sb/In BEP ratio of 27. In contrast to the small diameter variation of InAs/InSb heterostructure nanowires grown by MOVPE and CBE[3,4], here a diameter increase from 130% to 589% occurs. Based on previous works on InAs/InSb and GaAs/GaSb heterostructure nanowires, the following factors have been considered in evaluating the mechanism for the diameter increase. It is believed that the change in particle volume due to the uptake of group-III atoms is the main reason for the diameter increase[3,4,6-8]. A change of particle wetting angle/aspect ratio could also explain part of the diameter increase, but is believed to be only a minor effect[7]. Additionally, the effect of the rotation of the nanowire sidewalls was also considered[4]. We argue, however, that in our system the above reasons cannot be the critical factors for the diameter expansion since the particle volume is relative small, as shown in Table S1. We believe that the very large InSb segment diameter is mostly caused by non-seeded lateral growth on the side walls of the nanowires.

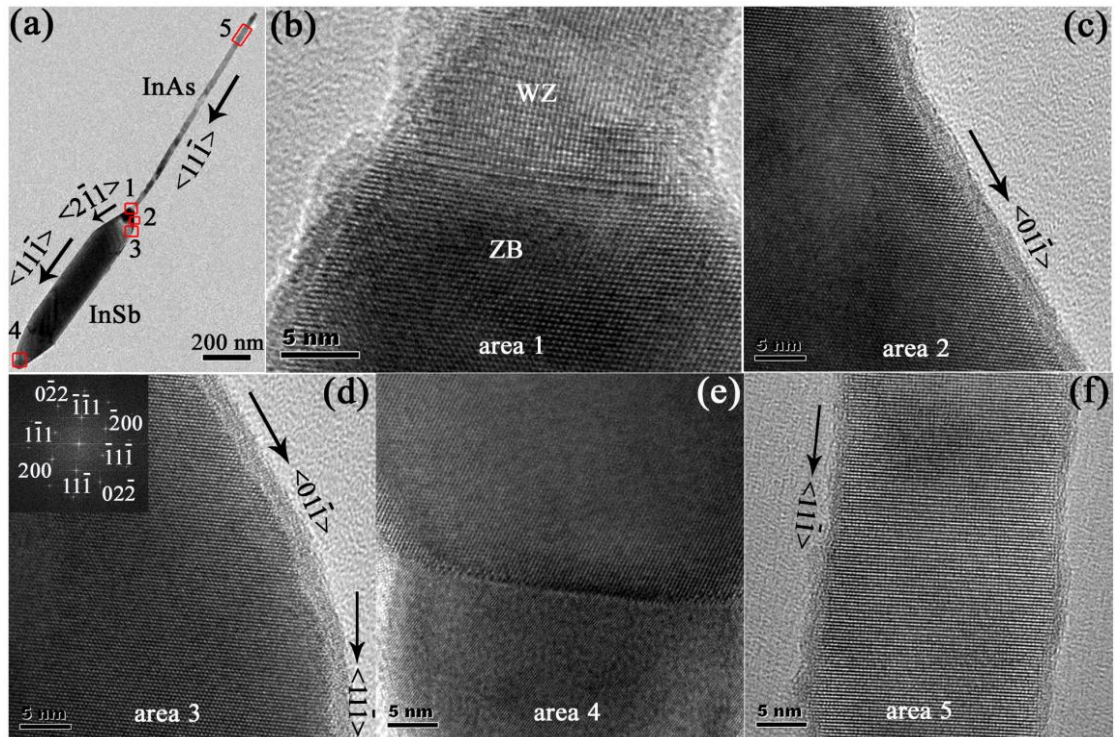

Fig. S3. (a) A low-resolution TEM image of an InAs/InSb axial heterostructure nanowire grown with an Sb/In BEP ratio of 27. The red rectangles highlight the regions where the high-resolution TEM images were recorded. (b-f) High-resolution TEM images of the heterostructure nanowire taken from regions 1, 2, 3, 4 and 5, respectively. The inset of (d) is FFT of the image. The black arrows denote the growth directions.

Figure S4 shows a top-view SEM image of InAs/InSb nanostructures grown with an Sb/In BEP ratio of 40, which showing the high yield of the InSb nanosheets. Each arrow indicates an InSb nanosheet. InSb nanosheets have different size which might be caused by different InSb local growth conditions such as the InAs nanowire diameter, length and growth direction *etc*.

| Sample Number | Diameter of InAs (nm) | Diameter of InSb (nm) | Diameter of catalyst particle (nm) | Increase of InSb diameter (%) | Increase of particle diameter (%) | Lengths for InAs/ InSb (μm) |
|---|---|---|---|---|---|---|
| 1 | 23 | 53 | 32 | 130 | 39.1 | 1.5/1.5 |
| 2 | 22 | 68 | 25 | 209 | 13.6 | 0.5/1.4 |
| 3 | 25 | 102 | 33 | 308 | 32.0 | 2.3/2.0 |
| 4 | 23 | 117 | 30 | 409 | 30.4 | 2.6/1.6 |
| 5 | 18 | 124 | 23 | 589 | 27.8 | 2.2/2.0 |
| 6 | 29 | 136 | 36 | 369 | 24.1 | 2.2/2.2 |

Table S1. List of parameters for six InAs/InSb axial heterostructure nanowires (grown with an Sb/In BEP ratio of 27) including diameters of the InAs and InSb sections and the catalyst particle, the percentage increase of the InSb and catalyst particle diameters and the lengths of InAs and InSb.

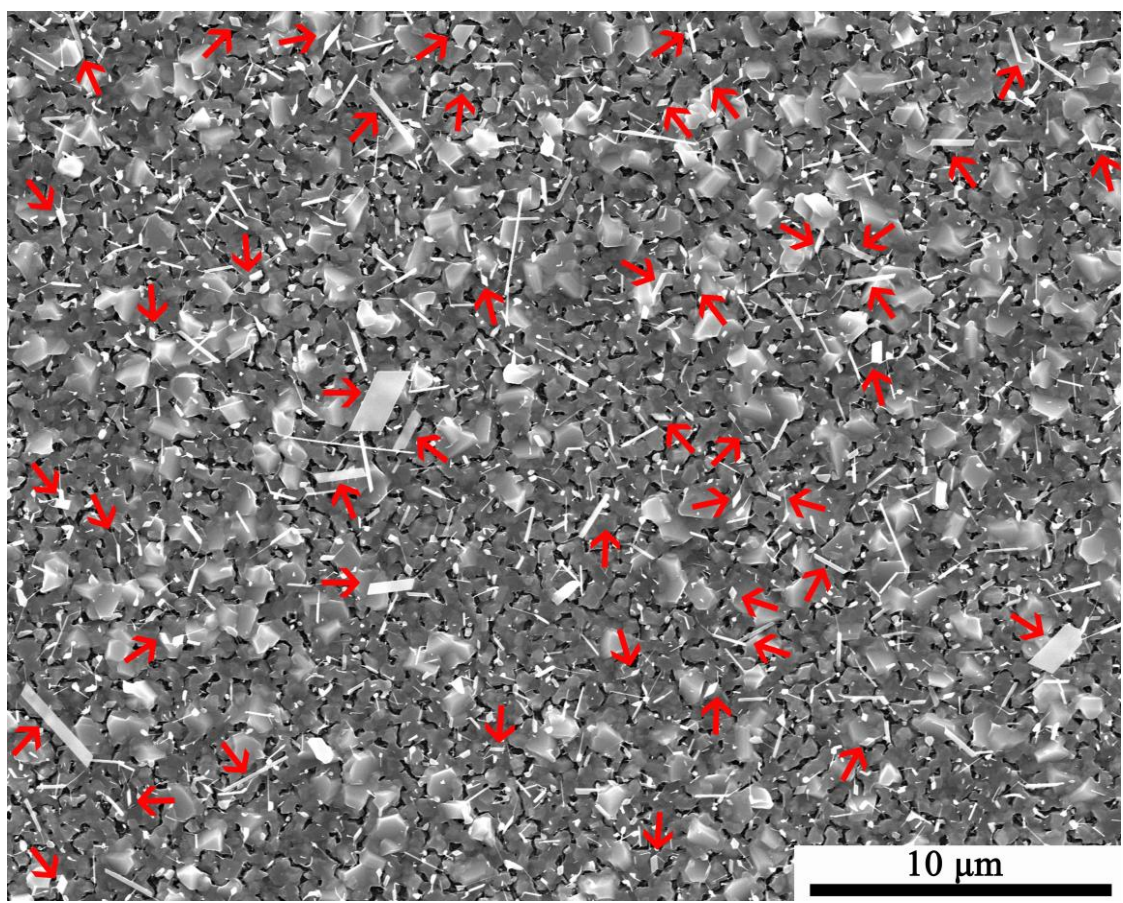

Fig. S4. A SEM image of InAs/InSb nanostructures grown with an Sb/In BEP ratio of 40.

## S2  Thickness information of the InSb nanosheets

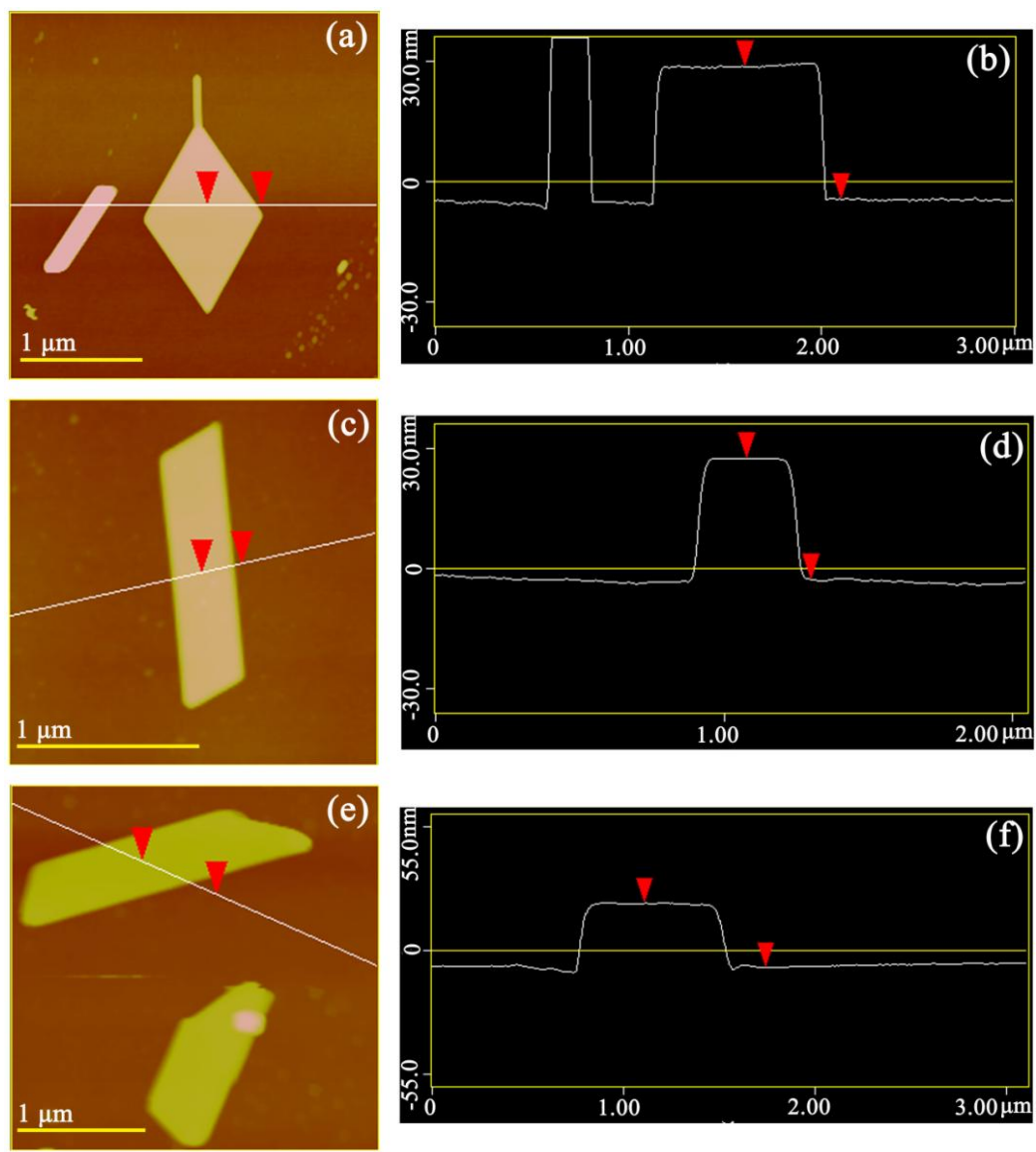

Fig. S5. (a), (c) and (e) AFM images of the InSb nanosheets with different sizes. (b), (d) and (f) Thickness information of the InSb nanosheets shown in (a), (c) and (e), respectively.

The thickness of the InSb nanosheets was measured using atomic force microscopy (AFM). Figures S5(a), S5(c) and S5(e) show the AFM images of the InSb nanosheets with different sizes. The length and width of the nanosheets shown in Figs. S5(a), S5(c) and S5(e) are 940 nm×940 nm (825 nm×217 nm, small one), 1400 nm×380 nm and 2800 nm×500 nm, respectively. As shown in Figs. S5(b), S5(d) and S5(f), the corresponding thicknesses of them are 33 nm (73 nm, small one), 30 nm and 28 nm, respectively.

## S3  Detailed crystal structure and quality information of InSb nanosheets

### S3.1 TEM images of an InSb nanosheet grown on a WZ InAs nanowire

We find that the epitaxial growth of InSb nanosheets on WZ InAs nanowires can also be realized. Figure S6 shows TEM images of an InSb nanosheet grown on a WZ InAs nanowire. Detailed high-resolution TEM images (Figs. S6(b) to S6(f)) indicate that the InSb nanosheet has a ZB crystal structure and no stacking faults or twinning defects are found in the corner and other sections of the nanosheet. The InAs nanowire has a pure WZ crystal structure with a diameter about 15 nm (Fig. S6(g)). Clearly, an atomically sharp structure interface (from WZ to ZB) can be observed at the InAs/InSb interface section (Fig. S6(c)).

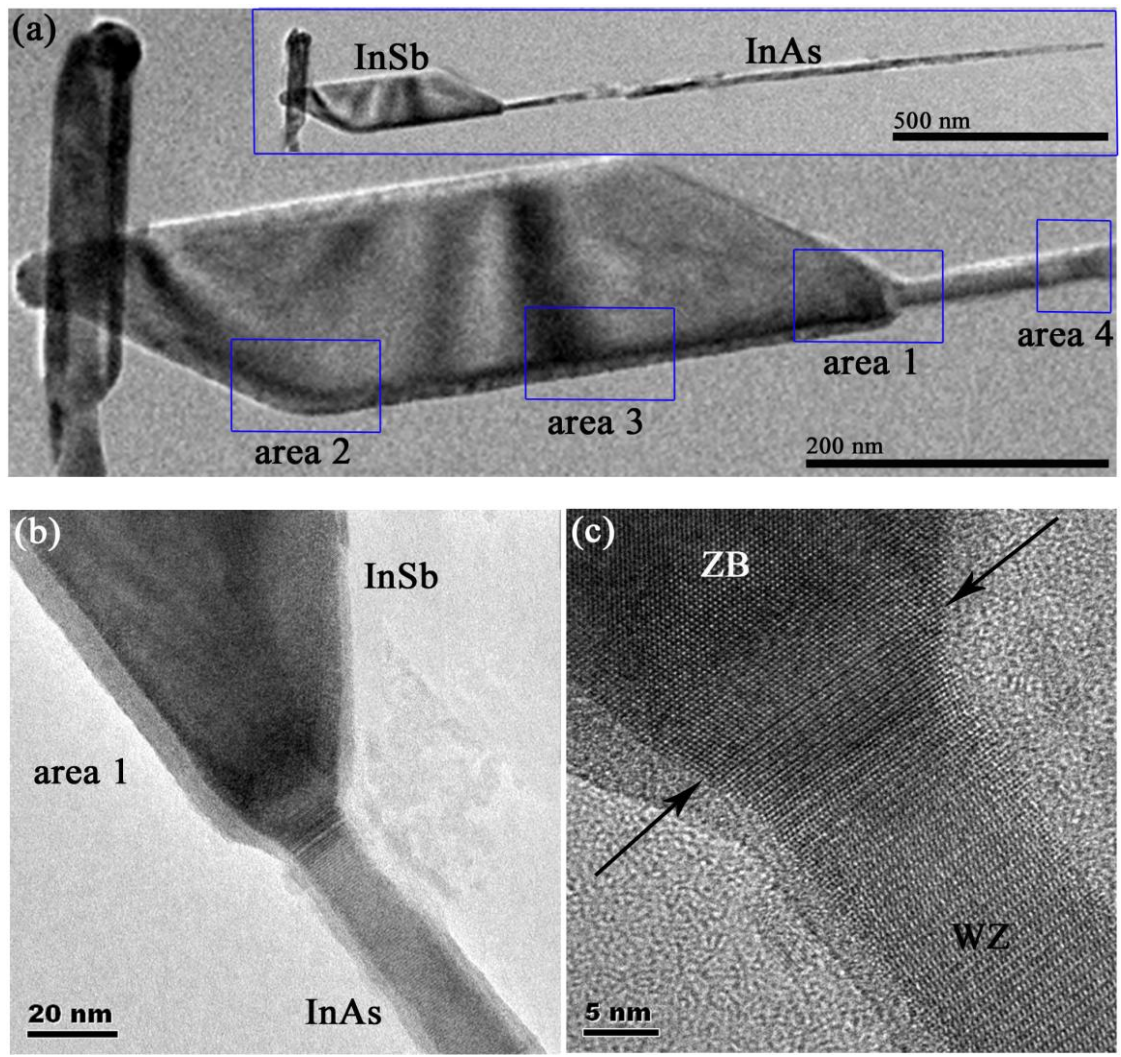

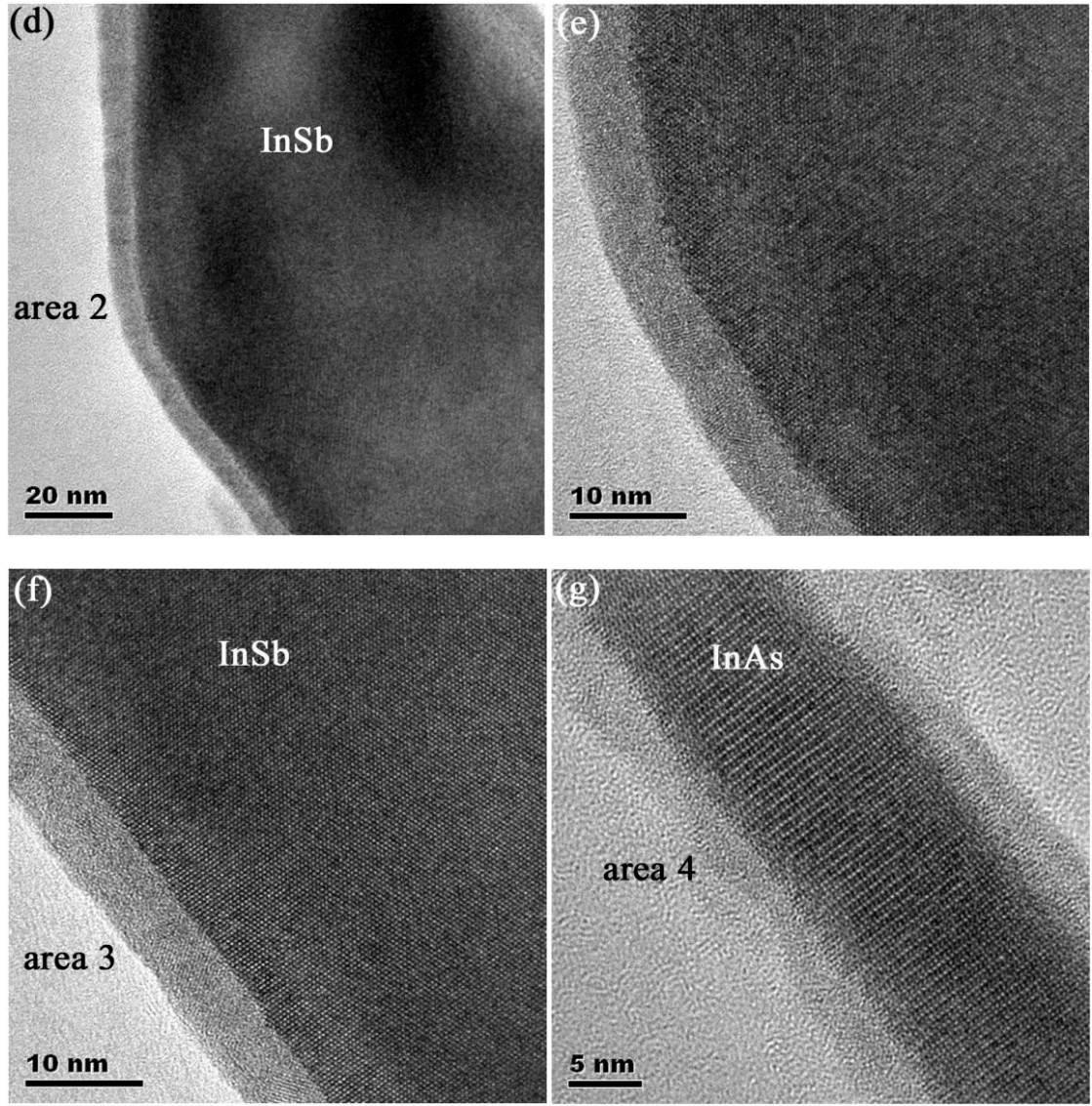

Fig. S6. (a) A bright-field TEM image of an InAs/InSb nanowire-nanosheet grown with an Sb/In BEP ratio of 27. Upper inset: the overall view of the nanowire-nanosheet; (b,c) High-resolution TEM images taken from area 1 (InAs/InSb interface section); (d,e) High-resolution TEM images taken from area 2 (corner section of InSb nanosheet); (f) High-resolution TEM images taken from area 3 (side section of InSb nanosheet); (g) High-resolution TEM images taken from area 4 (InAs nanowire section).

**S3.2 Selective area electron diffraction patterns of an InSb nanosheet**

To further examine the crystal quality of the InSb nanosheets, selective area electron diffraction pattern (SAED) were performed on different areas of the nanosheets. Figure S7 is a bright-field TEM image of a typical InAs/InSb nanowire-nanosheet. The InAs stem is 23 nm in diameter and 785 nm in length, while the InSb segment has a parallelogram shape with side lengths of 286 nm and 506 nm. Figures S7(a) to S7(e) are SAED patterns taken respectively near the tip, corner, middle and bottom of the InSb nanosheet as indicated by the red squares (a, b, c, d and e, respectively). The SAED patterns recorded from these spots of the nanosheet are identical and can be indexed to the face-centered cubic phase of InSb viewed along the [011] axis. The single set of diffraction spots further indicate that the nanosheet is a single crystal, free of twin defects and stacking faults. The growth directions of the InAs segment and the two sides of the InSb segment are $<\bar{2}1\bar{1}>$, $<0\bar{1}1>$ and $<1\bar{1}1>$, respectively, as shown in Fig. S7(f).

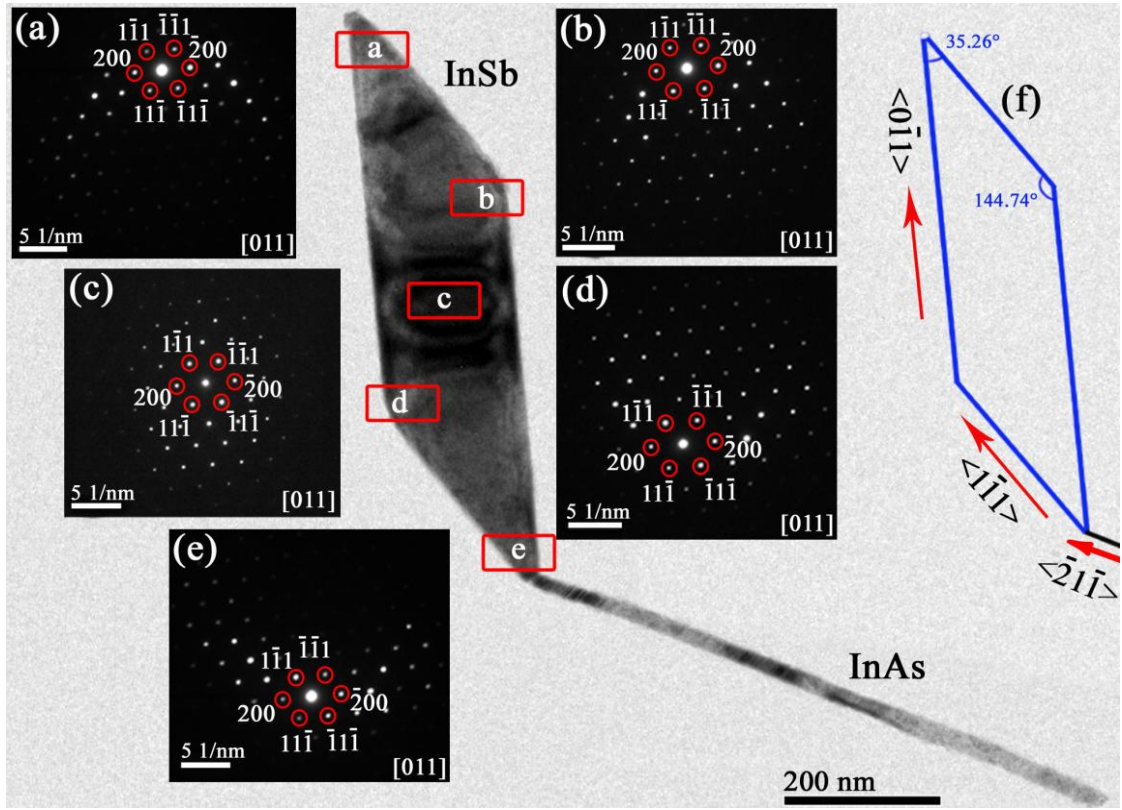

Fig. S7. A bright-field TEM image of an InAs/InSb nanowire-nanosheet grown with an Sb/In BEP ratio of 27. The red rectangles highlight the regions where SAED patterns were recorded. (a-e) SAED patterns taken along the [011] axis from regions a, b, c, d and e, respectively. (f) Schematic diagram illustrating the nanowire-nanosheet. The red arrows in (f) denote the growth directions of the InAs nanowire and sides of the InSb nanosheet.

**S3.3 TEM images of a large size InSb nanosheet**

To obtain the more detailed structure of the InSb nanosheets, high-resolution TEM investigations of several InSb nanosheets with large size were conducted and one set of results are shown in Fig. S8. Figure S8(a) is the bright-field TEM image of an InAs/InSb nanowire-nanosheet. The InSb segment has a parallelogram shape with side lengths of 370 nm and 1.4 μm. Figures S8(b), S8(d) and S8(f) show the high-resolution TEM images of the InSb section near the tip, the InAs/InSb interface, and the corner section of InSb, respectively; their corresponding magnified images are shown in Figs. S8(c), S8(e) and S8(g). As can be seen, no stacking faults or twinning is found in the corner and other sections of the InSb nanosheet, which is consistent with the SAED results above. The FFT patterns further show that the InSb nanosheet has a ZB crystal structure. It should be noted that at the upper part of the nanosheet a sharp interface between InSb and the catalyst particle is clearly visible. The FFT confirms that the particle on top of the InSb segment is single crystalline, with a hexagon crystal structure (Fig. S10). High-resolution TEM images together with their FFTs taken from the two sides of the InSb nanosheet (Figs. S8(h) and S8(i)) indicate that the two sides of the nanosheet are along the $<01\bar{1}>$ and $<\bar{1}1\bar{1}>$ directions, as illustrated in Fig. S8(j). Larger size InSb nanosheets (length and width are both several micrometers) with parallelogram, pentagon and hexagon shapes were also measured by TEM (not shown here), all of them have pure ZB crystal structures, completely free of stacking faults and twinning defects.

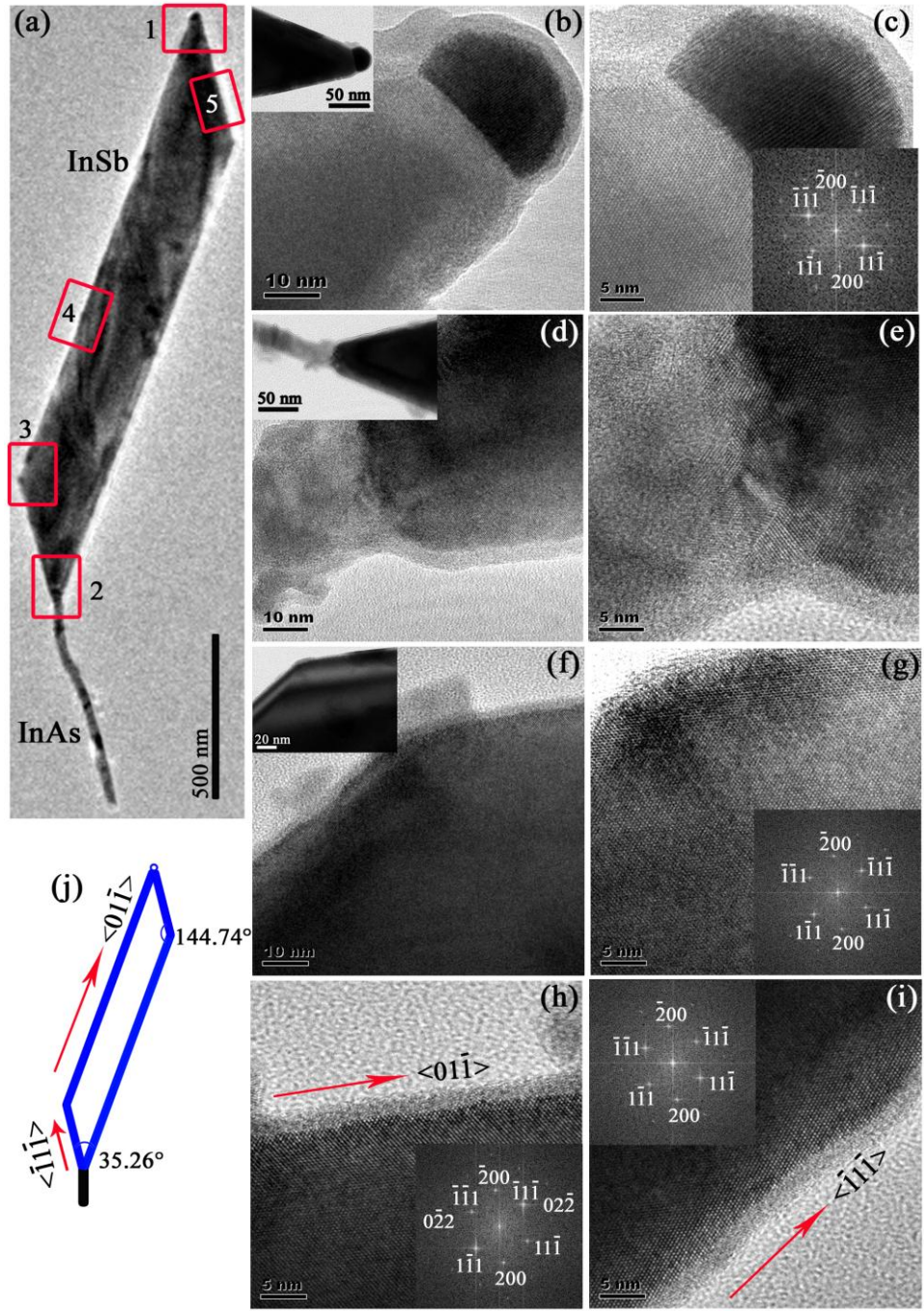

Fig. S8. (a) A bright-field TEM image of an InAs/InSb nanowire-nanosheet grown with an Sb/In BEP ratio of 27. The red rectangles highlight the regions where the high-resolution TEM images were recorded. (b and c), (d and e) and (f and g) high-resolution TEM images taken from regions 1, 2 and 3, respectively. The Insets of (b), (d) and (f) show magnified view of the region 1, 2 and 3, respectively. (h) and (i) High-resolution TEM images taken from the two sides marked with 4 and 5 of the InSb nanosheet, respectively. The insets in (c), (g), (h) and (i) are FFTs of the images. (j) Schematic diagram illustrating the nanowire-nanosheet. The red arrows in (h), (i) and (j) denote the growth directions of the sides of the InSb nanosheet.

## S4 Chemical composition of the InSb nanosheets

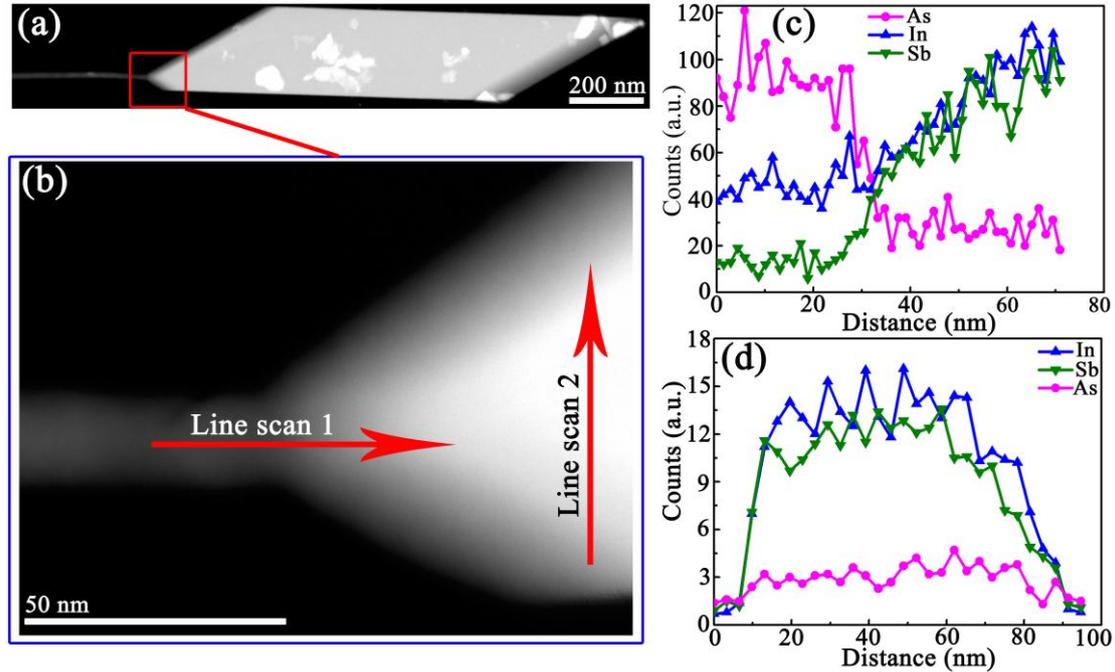

Fig. S9. (a) STEM image of an InAs/InSb nanowire-nanosheet. (b) The magnified view of the junction region (marked with red rectangle in (a). The red arrow highlights the directions where the EDS line scans were recorded. (c) and (d) EDS line scans taken in the InAs/InSb interface region (line scan 1 in (b)) and the InSb nanosheet region (line scan 2 in (b)), respectively.

The chemical composition of the InSb nanosheets was investigated by energy-dispersive x-ray spectrometer (EDS) analysis using both line scans and point analysis. On the EDS line scan of the InAs/InSb interface, it is observed that the heterostructure starts as InAs and then changes to InSb (Fig. S9(c)). The gradient observed for the Sb signal in Fig. S9(c), (this phenomenon is also observed in our InAs/InSb nanowires) is believed to result from the limited spatial resolution of the EDS spectra due to scattering effects in combination with a small overgrowth of InSb on the InAs part closest to the InSb segment as reported in the literature[3]. For comparison, the chemical composition of InAs/InSb heterstructure nanowires was also measured. Detailed EDS point analysis indicates that the InSb sections in both nanowires and nanosheets contain In and Sb with an atomic ratio of ~1:1 (Table S2). A small amount of As (less than 3%) was detected inside the InSb near the InAs/InSb interface by quantitative EDS point analysis and line scan (Figs. S9(c) and S9(d)). We note that in earlier studies[3,7-8], a small background of As was also detected. The As concentration is somewhat higher at the beginning of the InSb segment due to uptake of As from the stem but decrease rapidly in the InSb segment away from the heterojunction. The remaining particles for InSb nanowires and nanosheets are found to be composed of Ag, In and Sb, confirming the activity of Ag-In-Sb ternary alloys in the catalyzed growth[5,9].

| Composition (in at. %) | Sample Number | In | Sb | | Ag | In | Sb |
|---|---|---|---|---|---|---|---|
| InSb nanowires | 1 | 50.4 | 49.6 | catalyst particles | 61.4 | 26.1 | 12.5 |
| | 2 | 50.4 | 49.6 | | 62.8 | 25.2 | 11.4 |
| | 3 | 51.9 | 48.1 | | 65.3 | 24.5 | 10.1 |
| InSb nanosheets | 1 | 50.9 | 49.1 | catalyst particles | 61.0 | 26.0 | 13.0 |
| | 2 | 51.7 | 48.3 | | 63.2 | 26.0 | 10.8 |
| | 3 | 50.1 | 49.9 | | 61.8 | 25.0 | 13.2 |

Table S2. List of EDS point analysis of InSb nanowires, nanosheets and their catalyst particles (grown with an Sb/In BEP ratio of 27).

## S5  Seed-particles information of the InSb nanostructures

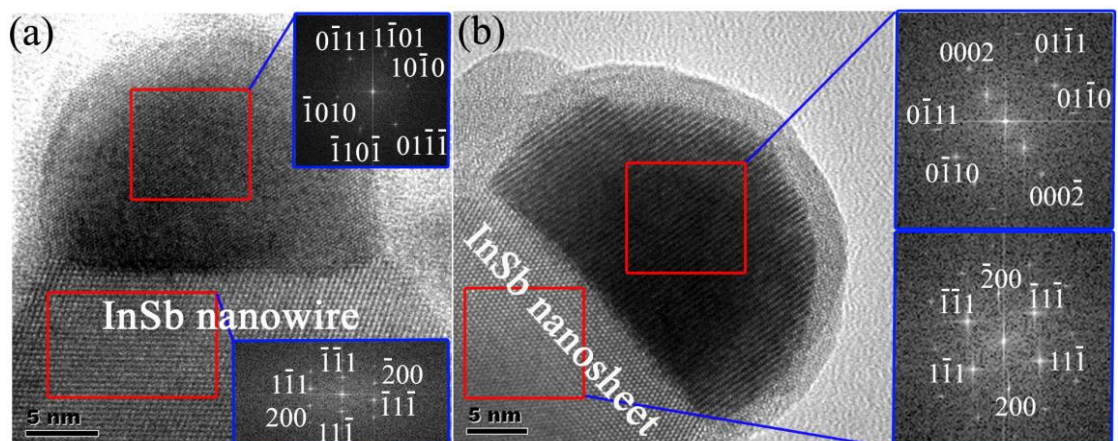

Fig. S10. (a) and (b) High-resolution TEM images of the InSb/seed-particle region taken from an InSb nanowire and an InSb nanosheet, respectively. Insets: the corresponding FFTs of the images.

The seed-particle crystal structure of the InSb nanowire and nanosheet has been compared. As shown in Figs. S10(a) and S10(b), high-resolution TEM studies indicate that the spherical catalyst particles for the InSb nanowire and nanosheet are both single-crystalline with the same hexagon crystal structures, which is different from Au-catalyzed InAs/InSb heterostructure nanowires[3,8] and InSb nanowires[10] with the ZB structure of the particles. As mentioned above, post-growth EDS analysis indicates that the contents of Ag, In and Sb in the post-growth seed particles of the InSb nanosheets and the InSb nanowires are similar (Table S2).

## S6  A nucleation process of the InSb nanosheets

Figures S11(a) to S11(g) show one vapor-liquid-solid (VLS) and anisotropic lateral growth process of InSb nanosheets. The VLS growth and the anisotropic lateral growth should take place at the same time (Fig. S11(b)) instead of sequentially as, evidenced by the fact that the nanosheets are very uniform in size, without any tapering and their sizes depend on the growth time as mentioned above. In our work, InSb evolves from thin nanowires to thick nanowires and to nanosheets with only increasing the Sb/In BEP ratio, we conjecture that the very high Sb flux could result in the formation of unstable droplets as reported in literatures[11-13]. The unstable droplets could crawl along the direction of triple phase line (one example shown in Figs. S11(b) to S11(g)), which promotes the growth of the InSb nanosheets with different shapes.

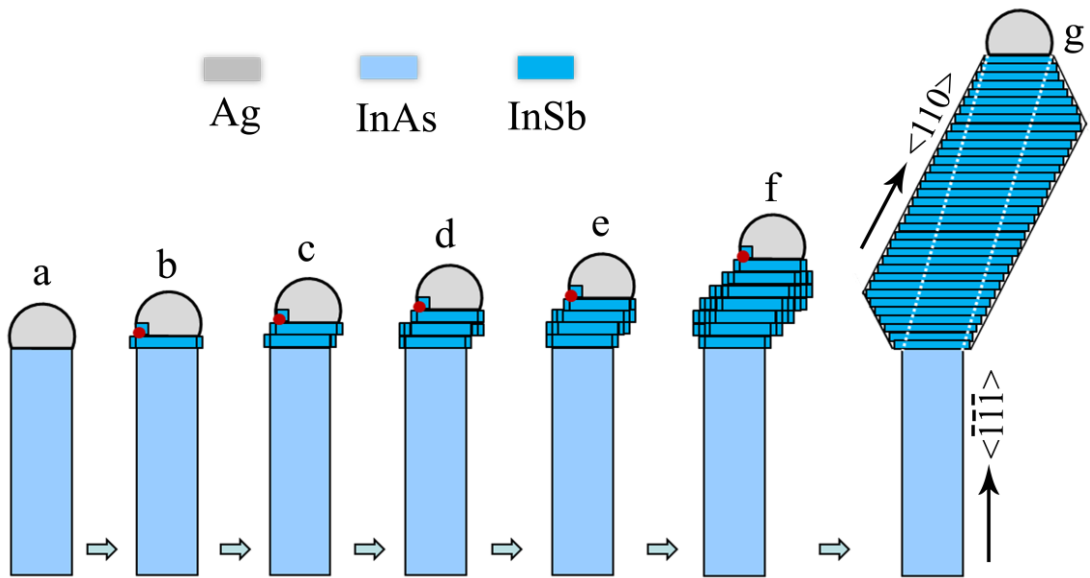

Fig. S11. Schematic demonstration of one nucleation process of the InSb nanosheets and the red dots indicate the positions of the triple phase line: (a) An InAs nanowire with a spherical droplet; (b) The growth of InSb combined axial VLS nucleation and anisotropic lateral growth, as well as the very high Sb flux, result in an unstable of the droplet; (c to f) The droplet crawls from left to right (or right to left, not shown here) along the direction of triple phase line and InSb continues growth in axial and lateral directions simultaneously; (g) Final result of an InSb nanosheet.

## S7  Summary of the Hall-bar device parameters

Two Hall-bar devices (Dev-1&2) were measured in detail in this work. Figure S12 shows the AFM images of Dev-1 and Dev-2. Table S3 summarizes the device parameters, including the Hall bar's length ($L$), width ($W$), nanosheet thickness ($T$), gate-to-nanosheet capacitance calculated from geometric size ($C_{g1}$) and from Hall measurement as discussed in the main text ($C_{g2}$). We note that after the sample is loaded into the dilution fridge chamber, a long, continuous vacuum pumping for more than 48 hours before cooling down is found to effectively reduce the devices' resistances by 1 order. Similar effect has been reported in MBE-grown InAs nanowires which may relate to surface de-adsorption of water molecules under high vacuum[14].

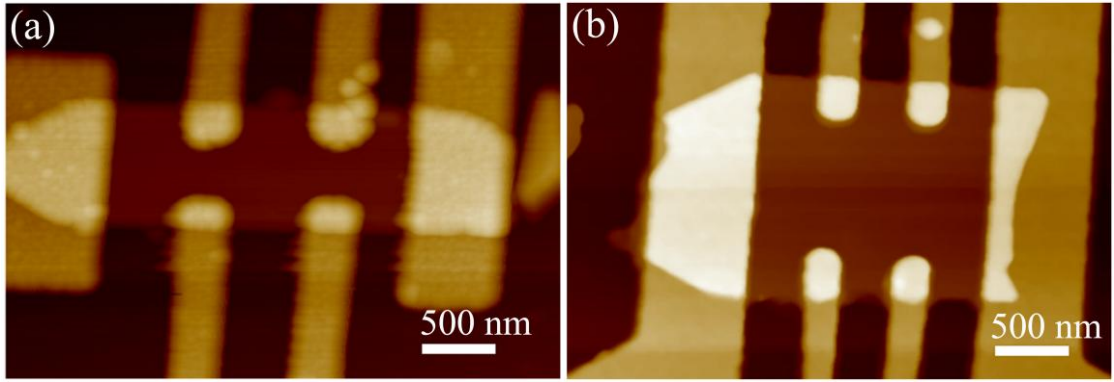

Fig. S12. AFM images of the two Hall-bar devices in this work. (a) Dev-1. (b) Dev-2. Part of the nanosheet in Dev-2 was broken off during the dry transfer process but does not show impact on the electrical measurement data.

| Dev # | $L$/nm | $W$/nm | $T$/nm | $C_{g1}$/μF | $C_{g2}$/μF |
|---|---|---|---|---|---|
| 1 | 2300 | 780 | 50 | 590 | 500 |
| 2 | 1642 | 1480 | 55 | 800 | 832 |

Table S3. Summary of device parameters. The two nanosheets have similar thickness of ~50 nm. The length-to-width ratio ($L/W$) is ~3 for Dev-1 and ~1.1 for Dev-2. Compared to the geometric capacitance ($C_{g1}$), the capacitance extracted from Hall measurement ($C_{g2}$) is ~4% larger for Dev-1, while is ~15% smaller for Dev-2.

## S8  Extraction of the field-effect mobility in a nanosheet

In a 2D planar transistor, the S/D current $I_{ds}$ is related to the total charge $Q$ via the equation

$$I_{ds} = |v_d| \frac{Q}{L} \qquad (1)$$

where $v_d$ is the carrier drift velocity and $L$ is the channel length.

In equation (1), $v_d$ and $Q$ can be substituted with the relation:

$$\vec{v_d} = \mu \vec{E} \qquad (2)$$

$$Q = C_g \left( V_{gs} - V_{th} \right) \qquad (3)$$

where $\mu$ is the carrier mobility (cm$^2$/Vs), $V_{th}$ is the threshold voltage, $\vec{E}$ is the electric field along the drift direction, and $C_g$ is the gate-to-channel capacitance. $\vec{E}$ has the following form

$$|\vec{E}| = \frac{V_{ds}}{L}, \qquad (4)$$

where $V_{ds}$ is the S/D bias.

Combining the upper expressions, we get the relation of $I_{ds}$ with gate voltage $V_{gs}$,

$$I_{ds} = \mu \frac{V_{ds}}{L^2} C_g \left( V_{gs} - V_{th} \right) \qquad (5)$$

The nanosheet conductance $G_s = I_{ds} / V_{ds}$, so we have

$$G_s = \frac{\mu}{L^2} C_g \left( V_{gs} - V_{th} \right) \qquad (6)$$

The derivative of $G$ to $V_{gs}$ is the transconductance $g$:

$$g = \frac{dG_s}{dV_{gs}} = \mu \frac{C_g}{L^2} \qquad (7)$$

Note that the measured resistance comprises of two terms: the sample resistance $G_s^{-1}$ and the contact resistance $R_c$,

$$G^{-1} = G_s^{-1} + R_c \qquad (8)$$

Equations (6) and (8) can be used to fit the on-state $G-V_{gs}$ curve with mobility $\mu$, contact resistance $R_c$ and threshold voltage $V_{th}$ as fitting parameters. This is the pinch-off trace method used in the main text to extract carrier mobility.

**S8.1 Hall measurement on Dev-1**

In the main text Fig. 5, we have shown the Hall results of Dev-2, which is used to extract the gate-to-nanosheet capacitance $C_g$. Similar measurements were also performed on Dev-1 and the data were shown in Fig. S13.

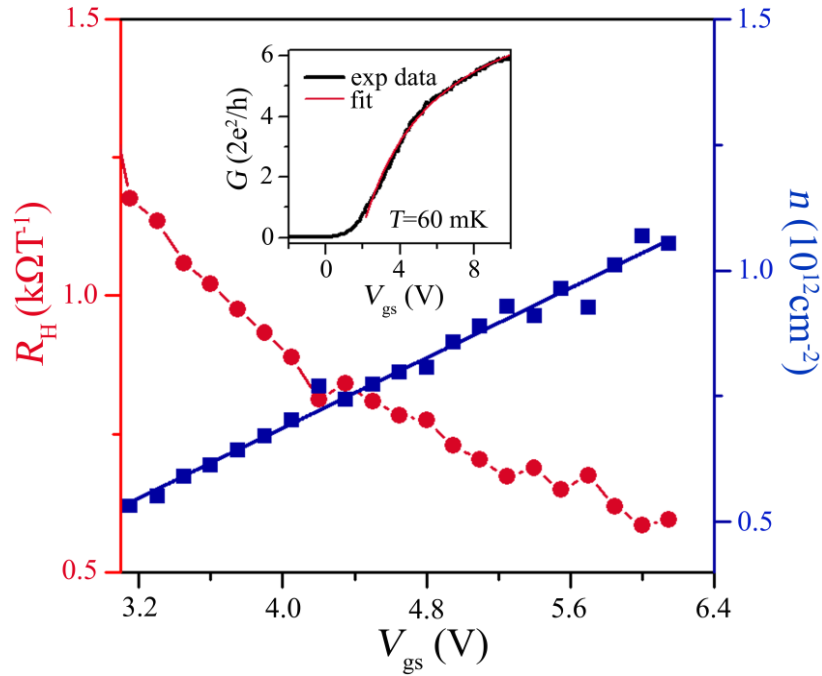

Fig. S13. Hall measurement and field-effect mobility fit of Dev-1. The main panel shows the extracted low-field Hall coefficients $R_H$ (red circle) and carrier densities $n$ (blue square) as a function of the back gate voltage $V_{gs}$. A linear fit to the slope of $n$-$V_{gs}$ yields a gate-to-nanosheet capacitance $C_g$ ~500 µF. The inset shows the field-effect mobility fit of Dev-1 using the same $C_g$ value at $T = 60$ mK. This fit gives a field-effect mobility estimate of ~18,500 cm$^2$/Vs.

**S8.2 Temperature-dependent field-effect motilities of Dev-1 and Dev-2**

Figure S14 shows Dev-1's field-effect mobility extracted as a function of temperature, corresponding to the $G-V_{gs}$ relation in the main text Fig. 4. Figure S15 shows a similar measurement on Dev-2.

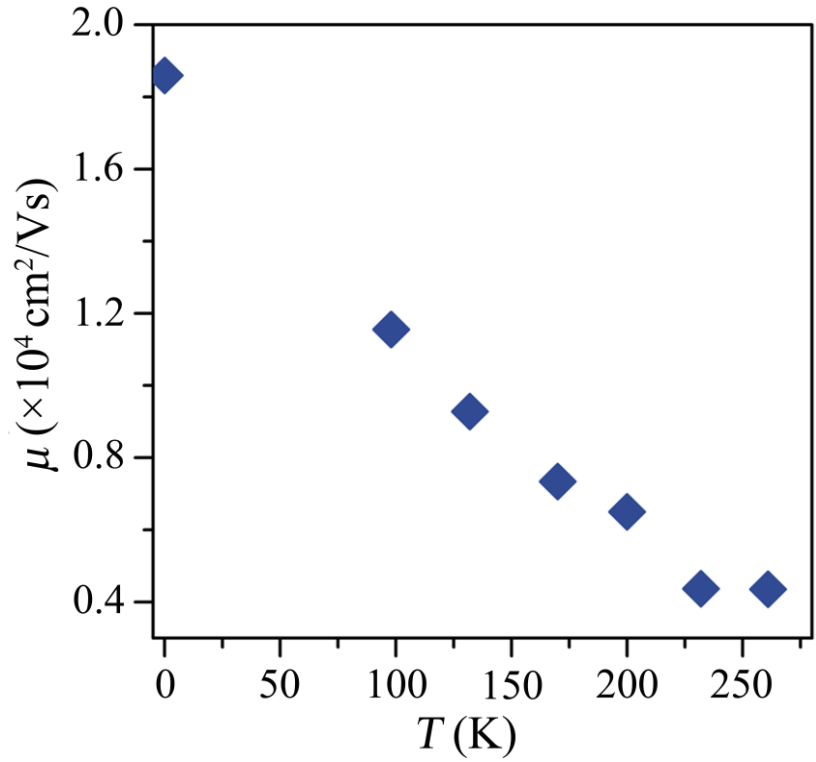

Fig. S14. The extracted field-effect mobility as a function of temperature for Dev-1. The mobility shows an increasing trend from ~4,000 cm$^2$/Vs at $T$ ~250 K with lowering temperature and has no sign of saturation in the low-$T$ region.

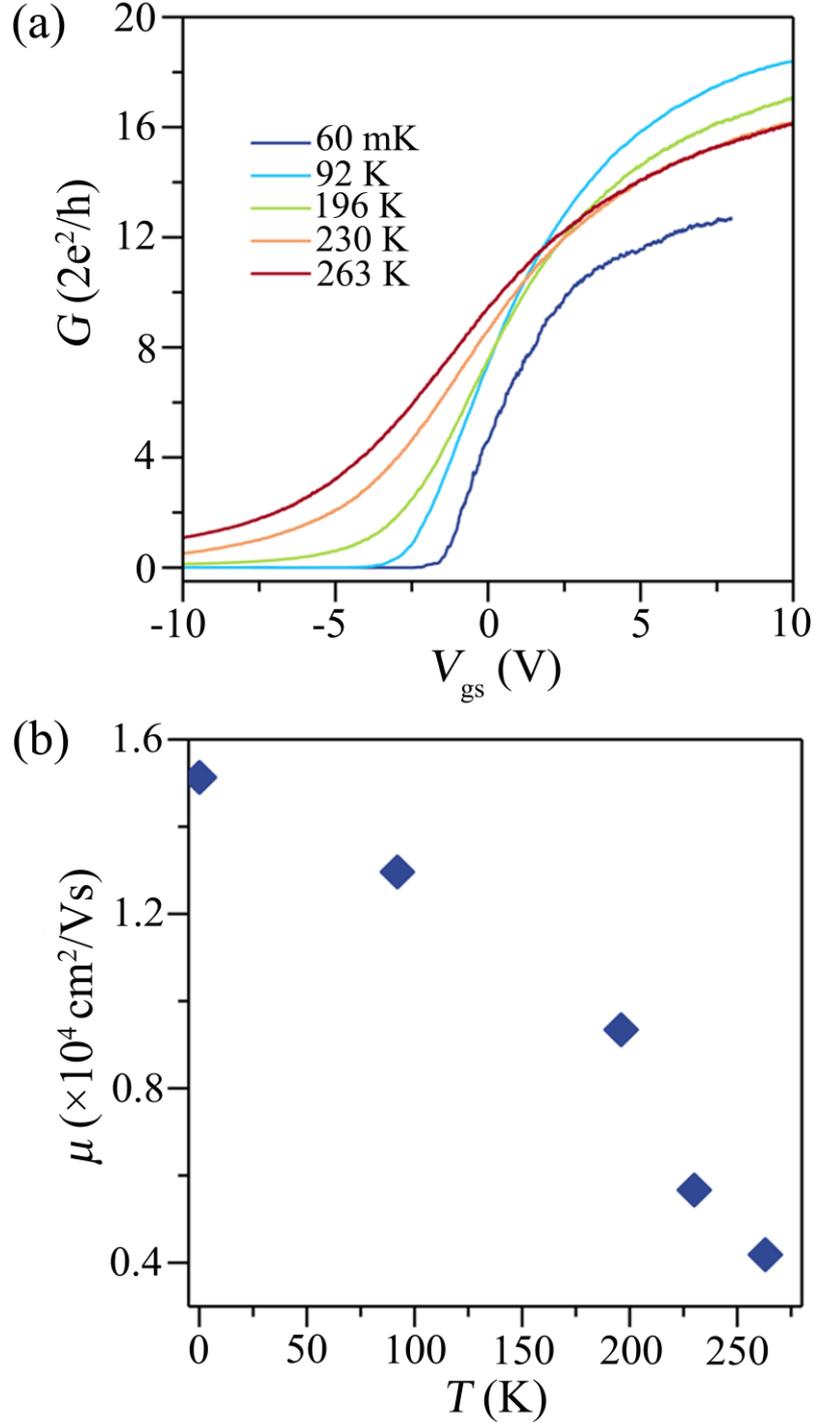

Fig. S15. Temperature-dependent transfer characteristics measured on Dev-2. (a) 2-probe $G$-$V_{gs}$ curves at various temperatures. Starting from 92 K, the on (off)-state conductance shows a decrease (increase) with increasing $T$, but the $G$-$V_{gs}$ curve at 60 mK has a much lower on-state conductance compared to the 92 K curve. This may relate to electron-electron interaction and phase-coherent transport processes such as universal conductance fluctuations at low temperatures, which is supported by the more prominent fluctuations seen in the $G$-$V_{gs}$ curve at 60 mK. (b) Extracted field-effect mobility of Dev-2 as a function of $T$. Similar mobility values ~4,000 cm$^2$/Vs are found for the two devices at $T$ > 250 K, but a slower increase is found for Dev-2.

# References

[1] Borg, B. M., Wernersson, L. E. Synthesis and properties of antimonide nanowires. *Nanotechnology* **24,** 202001 (2013).

[2] Debnath, M. C., Zhang, T., Roberts, C., Cohen, L. F., Stradling, R. A. High-mobility InSb thin films on GaAs (001) substrate grown by the two-step growth process. *J. Cryst. Growth* **267,** 17 (2004).

[3] Caroff, P. *et al*. High-quality InAs/InSb nanowire heterostructures grown by metal-organic vapor-phase epitaxy, *Small* **4,** 878 (2008).

[4] Lugani, L. *et al*. Faceting of InAs-InSb heterostructured nanowires. *Cryst. Growth Design* **10,** 4038 (2010).

[5] Yang, X. Y., Wang, G. M., Slattery, P., Zhang, J. Z., Li, Y. Ultrasmall single-crystal indium antimonide nanowires. *Cryst. Growth Design* **10,** 2479 (2010).

[6] Ghalamestani, S. G. *et al*. Demonstration of defect-free and composition tunable $Ga_xIn_{1-x}Sb$ nanowires. *Nano Lett.* **12,** 4914 (2012).

[7] Caroff, P. *et al*. InSb heterostructure nanowires: MOVPE growth under extreme lattice mismatch. *Nanotechnology* **20,** 495606 (2009).

[8] Ercolani, D. *et al*. InAs/InSb nanowire heterostructures grown by chemical beam epitaxy. *Nanotechnology* **20,** 505605 (2009).

[9] Park, H. D., Prokes, S. M., Twigg, M. E., Ding, Y., Wang, Z. L. Growth of high quality, epitaxial InSb nanowires. *J. Cryst. Growth* **304,** 399 (2007).

[10] Vogel, L. T. *et al*. Fabrication of high-quality InSb nanowire arrays by chemical beam epitaxy. *Cryst. Growth Design* **11,** 1896 (2011).

[11] Schwarz, K. W., Tersoff, J. Elementary processes in nanowire growth. *Nano Lett.* **11,** 316 (2011).

[12] Musin, I. R., Filler, M. A. Chemical Control of Semiconductor nanowire kinking and superstructure. *Nano Lett.* **12,** 3363 (2012).

[13] Tian, B., Xie, P., Kempa, T. J., Bell, D. C., Lieber, C. M. Single-crystalline kinked semiconductor nanowire superstructures. *Nature Nanotechnol.* **4,** 824 (2009).

[14] Kretinin, A. V., Popovitz-Biro, R., Mahalu, D., Shtrikman, H. Multimode Fabry-Pérot conductance oscillations in suspended stacking-faults-free InAs nanowires. *Nano Lett.* **10,** 3439 (2010).